\documentclass[manuscript,screen]{acmart}
\AtBeginDocument{%
  \providecommand\BibTeX{{%
    \normalfont B\kern-0.5em{\scshape i\kern-0.25em b}\kern-0.8em\TeX}}}

\setcopyright{acmlicensed}
\copyrightyear{2023}
\acmYear{2023}
\acmDOI{XXXXXXX.XXXXXXX}

\acmJournal{TODAES}
\acmVolume{37}
\acmNumber{4}
\acmArticle{111}
\acmMonth{8}

\acmPrice{15.00}
\acmISBN{978-1-4503-XXXX-X/18/06}

\usepackage[utf8]{inputenc}
\usepackage[most]{tcolorbox}

\usepackage{booktabs}

\usepackage{graphicx}
\usepackage{multirow}
\usepackage[noadjust]{cite} 
\usepackage{setspace}
\usepackage{longtable}
\usepackage{amsmath}
\usepackage{amsfonts} %
\usepackage{calligra} %
\usepackage{mathtools}
\usepackage{amsthm} % for neater definitions
\usepackage{pgfplots}
\usepackage{pgfplotstable}
\usepgfplotslibrary{statistics}
\makeatletter
\let\MYcaption\@makecaption
\makeatother
\usepackage[font=footnotesize]{subcaption}
\makeatletter
\let\@makecaption\MYcaption
\makeatother
\usepackage{comment}
\usepackage{listings}
\usepackage{tcolorbox}
\usepackage{circledsteps}
\usetikzlibrary{patterns}

\usepackage{blindtext}
\usepackage{etoolbox}
\usepackage{acronym}
\usepackage{multicol}
\usepackage{xcolor, colortbl}
\usepackage{tabularx}
\usepackage[most]{tcolorbox}

\usepackage{pifont}
\usepackage{float}
\usepackage{textcomp}
\definecolor{linkcolor}{RGB}{219, 48, 122}

\usepackage{forest}
\newenvironment{review}
    {\par\color{blue}}
    {\color{black}\par}
\renewenvironment{review}{}{}

\usepackage{algorithm}
\usepackage{algorithmic}
\usepackage{url}
\usepackage[most]{tcolorbox}
\usepackage{varwidth}
\tcbuselibrary{xparse}
\tcbuselibrary{skins}
\usepackage[all=normal, floats=tight,indent=tight]{savetrees}
\usepackage{pifont}
\usepackage{framed}
\usepackage{soul}
\usepackage{multirow}
\usepackage{array}
\newcolumntype{L}[1]{>{\raggedright\let\newline\\\arraybackslash\hspace{0pt}}m{#1}}
\newcolumntype{C}[1]{>{\centering\let\newline\\\arraybackslash\hspace{0pt}}m{#1}}
\newcolumntype{R}[1]{>{\raggedleft\let\newline\\\arraybackslash\hspace{0pt}}m{#1}}
\newcolumntype{H}{>{\collectcell\lstinline}l<{\endcollectcell}}
\usepackage{url}
\newcommand{\rt}[1]{{\color{red}{#1}}}
\renewcommand{\rt}[1]{#1}

\newcommand*\circled[1]{\tikz[baseline=(char.base)]{
            \node[shape=circle,draw,inner sep=0.7pt] (char) {#1};}}
            
\usepackage[T1]{fontenc}
\usepackage{hyperref}

\input{section/preamble/preamblelistingsconf}
\acrodef{CPS}{Cyber-Physical System}
\acrodef{IoT}{Internet of Things}
\acrodef{HDL}{hardware description language}
\acrodef{CAD}{Computer-Aided Design}
\acrodef{EDA}{Electronic Design Automation}
\acrodef{HPC}{High-Performance Computing}
\acrodef{DL}{deep learning}
\acrodef{ML}{machine learning}
\acrodef{NLP}{natural language processing}
\acrodef{IC}{Integrated Circuit}
\acrodef{CWE}[CWE]{Common Weakness Enumeration}
\acrodef{CVE}[CVE]{Common Vulnerabilities and Exposures}
\acrodef{LLM}[LLM]{large language model}
\acrodef{NMT}[NMT]{neural machine translation}
\acrodef{NLP}[NLP]{natural language processing}
\newcommand{\ignore}[1]{{}}

\newcommand{\squishlist}{
	\begin{list}{$\bullet$}
		{ \setlength{\itemsep}{0pt}
			\setlength{\parsep}{1pt}
			\setlength{\topsep}{1pt}
			\setlength{\partopsep}{0pt}
			\setlength{\leftmargin}{0.9em}
			\setlength{\labelwidth}{1.5em}
			\setlength{\labelsep}{0.4em} } }
	\newcommand{\squishend}{
	\end{list}  }

\definecolor{graphFirst}{RGB}{2,136,209} % Light Blue 700
\definecolor{graphSecond}{RGB}{211,47,47} % Red 700
\definecolor{graphThird}{RGB}{245,124,0} % Orange 700
\definecolor{graphFourth}{RGB}{56,142,60} % Green 700
\definecolor{graphFifth}{RGB}{81,45,168} % Deep Purple 700
\definecolor{graphSixth}{RGB}{69,90,100} % Blue Grey 700
\definecolor{graphSeventh}{RGB}{251,192,45} % Yellow 700\definecolor{backgroundFirst}{RGB}{129,212,250} % Light Blue 200
\definecolor{backgroundSecond}{RGB}{239,154,154} % Red 200
\definecolor{backgroundThird}{RGB}{255,204,128} % Orange 200
\definecolor{backgroundFourth}{RGB}{165,214,167} % Green 200
\definecolor{backgroundFifth}{RGB}{179,157,219} % Deep Purple 200
\definecolor{backgroundSixth}{RGB}{176,190,197} % Blue Grey 200
\definecolor{backgroundSeventh}{RGB}{255,245,157} % Yellow 200

\begin{document}


% feel free to cahnge the title as long as it is not DAVE

\title{VeriGen: A Large Language Model for Verilog Code Generation}

\titlenote{This manuscript extends work presented at the Design, Automation \& Test in Europe Conference (DATE 2023)~\citep{thakur_benchmarking_2023}% S. Thakur et al., "Benchmarking Large Language Models for Automated Verilog RTL Code Generation," doi: 10.23919/DATE56975.2023.10137086.
}

% \author{Shailja Thakur}
\author{Shailja Thakur}
\email{st4920@nyu.edu}
\affiliation{%
  \institution{New York University}
  \city{New York}
  \country{USA}
}
\author{Baleegh Ahmad}
\email{ba1283@nyu.edu}
\affiliation{%
  \institution{New York University}
  \city{New York}
  \country{USA}
}
\author{Hammond Pearce}
\email{hammond.pearce@unsw.edu.au}
\affiliation{%
  \institution{University of New South Wales}
  \city{Sydney}
  \country{Australia}
}
\author{Benjamin Tan}
\email{benjamin.tan1@ucalgary.ca}
\affiliation{%
  \institution{University of Calgary}
  \city{Calgary}
  \country{Canada}
}

\author{Brendan Dolan-Gavitt}
\email{brendandg@nyu.edu}
\affiliation{%
  \institution{New York University}
  \city{New York}
  \country{USA}
}
 
\author{Ramesh Karri}
\affiliation{%
  \institution{New York University}
  \city{New York}
  \country{USA}
}
\email{rkarri@nyu.edu}

\author{Siddharth Garg}
\affiliation{%
  \institution{New York University}
  \city{New York}
  \country{USA}
}
\email{siddharth.garg@nyu.edu}

\authorsaddresses{This research work was supported in part by NSF Award 1553419, NSF Award 1646671, NSF Award 2039607, and ARO Award 77191NC. The opinions, findings, and conclusions, or recommendations expressed are those of the author(s) and do not necessarily reflect the views of any sponsors.
\\
Corresponding author: Shailja Thakur st4920@nyu.edu.
\\
Authors' addresses: Shailja Thakur, st4920@nyu.edu, New York University, New York, USA; Baleegh Ahmad, ba1283@nyu.edu, New York University,
New York, USA; Hammond Pearce, hp2265@nyu.edu, University of New South Wales, Sydney, Australia; Benjamin Tan, benjamin.tan1@ucalgary.ca, University of Calgary, Calgary, Canada; Brendan Dolan-Gavitt, brendandg@nyu.edu, New York University, New York, USA; Ramesh Karri, New York University, New York, USA, rkarri@nyu.edu; Siddharth Garg, New York University, New York, USA, siddharth.garg@nyu.edu}


\renewcommand{\shortauthors}{Thakur et al.}

\begin{abstract}
\rt{In this study, we explore the capability of Large Language Models (LLMs) to automate hardware design by generating high-quality Verilog code, a common language for designing and modeling digital systems. We fine-tune pre-existing LLMs on Verilog datasets compiled from GitHub and Verilog textbooks. We evaluate the functional correctness of the generated Verilog code using a specially designed test suite, featuring a custom problem set and testing benches. Here, our fine-tuned open-source CodeGen-16B model outperforms the commercial state-of-the-art GPT-3.5-turbo model with a 1.1\% overall increase.
Upon testing with a more diverse and complex problem set, we find that the fine-tuned model shows competitive performance against state-of-the-art gpt-3.5-turbo, excelling in certain scenarios. Notably, it demonstrates a 41\% improvement in generating syntactically correct Verilog code across various problem categories compared to its pre-trained counterpart, highlighting the potential of smaller, in-house LLMs in hardware design automation.}
% With a guided system prompt, GPT-3.5-turbo generates 40.6% correct code vs CodeGen-FT-16B 41.07%
% Further, when analyzing functional correctness, a fine-tuned open-source CodeGen LLM can outperform the state-of-the-art commercial Codex LLM (6.5\% overall). % and can construct error-free intermediate- and advanced-level Verilog hardware designs.
We release our training/evaluation scripts and LLM checkpoints as open-source contributions.

\end{abstract}


;



\maketitle

\section{Introduction}
\label{sec:intro}
% \todoblock{review the section}
%  \todoblock{Design productivity gap, success in software language}
% \todo{figure 1 to first page?}
Digital hardware design flows involve designers writing code in \acp{HDL} such as Verilog and VHDL to specify hardware architectures and behaviors, a process that is both time-consuming and bug-prone~\citep{dessouky_hardfails_2019}. 
%human-in-the-loop makes  the design extremely slow and buggy. 
%Attempts have been made to reduce human intervention in the design flow, such as OpenRoad, which leverage machine learning capabilities to automate design. [Unrelated]
As design complexity grows, there is a need to reduce design costs and developer effort during hardware specification.
\rt{Several attempts have thus sought to improve HDL design time and quality, for instance by using high-level synthesis---this allows developers to specify functionality in languages like C but comes at the expense of hardware efficiency. Related works also consider modernizing HDLs by adopting features and languages typically used in software development, for instance Chisel~\citep{bachrach_chisel_2012} based on Scala.}

\rt{A promising new approach comes via the proliferation of technically capable code-writing \acp{LLM}~\citep{chen_evaluating_2021}. 
\acp{LLM} are deep neural networks, typically based on transformer~\citep{vaswani_attention_2017} architectures, that aim to model the underlying distribution of a natural or structured language corpus. 
Given a sequence of words (or ``tokens'') \acp{LLM} predict a distribution over the next word/token.
When used in a loop (`autoregressively') with some `input prompt' and strategies for choosing the best tokens from the distribution, \acp{LLM} can thus complete text: If prompted with the first sentence of a paragraph of English prose, the model will suggest more text; or, if provided technical specifications or coding comments, LLMs can then implement the matching code.
The proficiency of these pre-trained models are a result of recent advance in scaling transformer architectures as well as the availability of massive text and code corpus from open-source repositories such as GitHub and StackOverflow.
%As such, developers can write specifications in plain language and then use the \acp{LLM} to generate HDL code directly.
As such, we wish to investigate the potential for developers to use \acp{LLM} to automatically generate the HDL code directly from a natural language like English.}
%to \emph{automatically} generate code from natural language specifications. 

\rt{While considerable effort has been undertaken which explores the training and utilization of language models for writing software~\citep{chen_evaluating_2021,ahmad_towards_2023,li_starcoder_2023} as well as in their down-stream effects (e.g. security~\citep{pearce_asleep_2022,sandoval_lost_2023}), there is scant examination of \acp{LLM} for hardware applications. As such, this was the focus of the original work~\citep{thakur_benchmarking_2023} as well as in this extension, where we perform the first comprehensive evaluation of the syntactic and functional correctness of synthesizable Verilog code generated by both open-source and commercial \acp{LLM}.}

%Recently, \acp{LLM} have been shown to successfully generate code in languages like C and Python~\citep{usenix23,manyothers}, \todo{broken refs?} but their use in generating HDL code has not been comprehensively evaluated. %has not been investigated and is the focus of this study.
%---paper studies whether this success can be translated for Verilog code generation.% directly from developer comments. 
%partially or even fully automate hardware specification by assisting human developers 
%and reduce the gap between the complexities of the hardware and the requisite design skills. % required to implement the design. 
%They can complete code from a comment and/or initial lines of code.
%assign probabilities to the sequence of words. Transformer-based large language models can generate code given an informal natural language description (\textit{prompt}) of the code.  

\rt{There are several key observations when considering the state of the art in code-writing LLMs for hardware. 
First, it has been observed that existing commercial LLMs, including GitHub Copilot, may generate Verilog code which fails syntax, synthesis, and functional checks~\citep{pearce_asleep_2022}. %\textcolor{blue}{(SG: Is this true? Do we show this?)} 
Secondly, while fine-tuning a given \ac{LLM} over a Verilog corpus can aid in its authorship of HDL~\citep{pearce_dave_2020}, this requires a large dataset of Verilog. Unfortunately, such datasets for Verilog are lacking, and so the prior work in this area used a small, synthetic template-generated HDL corpus meaning that the resultant model did not generalize to unseen problems.
Thirdly, quantifying model performance in this area is difficult: while datasets like HumanEval exist for software~\citep{chen_evaluating_2021}, large-scale test problems and methods to evaluate the syntactical and functional correctness of LLM-generated HDL are lacking.}

% While fine-tuning \acp{LLM} on a Verilog corpus can help, 
% it requires a large dataset of Verilog code which is lacking. 
% Prior work has used small, synthetic, template-generated HDL code corpus, but the resulting models do
% %trained models on template-generated hardware specifications and corresponding Verilog; this is time-consuming and does 
% not generalize to unseen problems~\citep{pearce_dave_2020}. 
% Finally, test problems and methods to evaluate the syntactical and functional correctness of LLM-generated code on a large scale are lacking. 
% %We finetune \acp{LLM} on Verilog from GitHub and textbooks. The Verilog corpus from textbooks benefits from a large number of Verilog textbooks, broader problem coverage, and verified solutions in the form of examples.

\begin{figure}[h]
    \centering
    \includegraphics[width=0.7\linewidth]{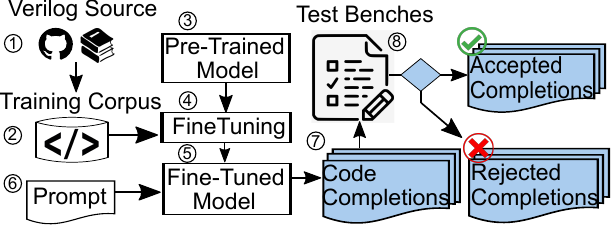}
    \caption{Experimental Evaluation of LLM Verilog Completions}
    \label{fig:system_overview}
\end{figure}

Our paper thus makes the following contributions. (1) By consolidating available open-source Verilog code we create (to the best of our knowledge) the largest training corpus of Verilog code yet used for training \acp{LLM}. 
(2) Using this corpus, we examine fine-tuning five different pre-trained \acp{LLM} models with parameter counts ranging from 345M to 16B, producing five new fine-tuned open-source models specialized for Verilog code generation.
(3) To evaluate the efficacy of the models and determine the effect of the parameter sizes, we design a set of Verilog coding problems with varying difficulty levels 
and corresponding test benches to test the functional correctness of generated code.
\rt{
(4) 
%We extend our prior work~\citep{thakur_benchmarking_2023} by performing a 
We comprehensively evaluate our fine-tuned models against state-of-the-art general-purpose \acp{LLM} including ChatGPT (GPT-3.5-turbo and GPT4) and PALM2 an expanded evaluation with a new set of Verilog design challenges. 
These \acp{LLM} are several times larger than our models, black-boxed and incur per-token costs; in contrast, our models are smaller, open-sourced, free, and provide comparable or improved performance. 
(5) We augment our training set with a selection of Verilog textbooks and provide detailed results on the performance of models tuned on this augmented dataset. 
% (6) We offer an expanded discussion on the current state-of-the-art following our evaluation, including perspectives on future work in this domain. 
% } \textcolor{blue}{(SG: 6 cannot be claimed as a new contribution in this section. It will turn off reviewers. We can list it separately in our letter highlighting addition to the DATE manuscript.)
}
To aid the community in performing further research in this area, we provide the training/evaluation code\footnote{\url{{https://github.com/shailja-thakur/VGen}}} and LLM checkpoints\footnote{\url{https://huggingface.co/shailja}}.
% Since prompt quality impacts the generated code~\citep{}, we design three prompts for each test problem with increasing levels of detail.

%We also design an evaluation pipeline that runs across all the problems and returns a pass or fail for each problem, depending on whether the generated code passes all the test cases.

\autoref{fig:system_overview} illustrates our experimental platform for studying the impact of parameters such as temperature, number of sequences generated per problem, and number of LLM parameters. 
\autoref{sec:data} discusses creating the training data from GitHub and PDFs of Verilog textbooks \circled{1} with pre-processing \circled{2} and the five pre-trained LLMs \circled{3} that we fine-tune \circled{4} for completing Verilog code \circled{5}. 
\autoref{sec:evaluation} explains the evaluation setup, including our hand-designed prompts \circled{6}. 
\autoref{sec:Results} presents our results from generating code suggestions \circled{7} and evaluating with an analysis pipeline that compiles the Verilog and checks it against unit tests \circled{8}. 
\autoref{sec:discussion} discusses how our evaluation shows that the largest code-based \acp{LLM} (i.e., CodeGen-16B) fine-tuned on our Verilog corpus demonstrate competitive performance over all other evaluated \acp{LLM}. Qualitatively, our best-performing fine-tuned \acp{LLM} can generate functioning code for challenging problems. 
%This hints that they generalize better on low-resource languages such as Verilog. 
%Our framework will be open-sourced to spur further research [link redacted for review].

% \todoblock{Consolidate and make into paper roadmap}

% \rt{
% \subsection{Overview of Experimental Method}
% \autoref{fig:system_overview} is a high-level overview of the system for synthesizing Verilog using LLMs. 

% Given a prompt \circled{6} (e.g., \autoref{fig:prompt-details})
% the fine-tuned LLM \circled{5} generates a set of \circled{7} code suggestions for the body section of the input prompt.

% For each prompt, the \ac{LLM} suggests several options for completion; each suggestion is a variant of the complete logic for the input problem. 
% To evaluate the quality of the completed code, we hand-design a set of problems, described in \autoref{tbl:problem_set}, and manually generate solutions and test benches for the problem. 

% Next, each of the completed problem variants is passed to \circled{8} the code analysis pipeline that compiles the Verilog, and checks it against the unit tests.
% A design that passes all the unit tests is accepted as a solution for the problem, otherwise, it is rejected. 
% }

% \clearpage

\section{Background and Related Work} \label{sec:background}
% \todoblock{talk about transformers and their applications}
% \todoblock{talk about DAVE, and other code completion techniques}
% \todoblock{combine the following model description into one paragraph, finally conclude saying why do we chose the varying set of models for evaluating our approach}
% DAVE~\citep{pearce_dave_2020}. Talk about Codex~\citep{openai_openai_2021}, Copilot~\citep{github_github_nodate}, GPT or Transformer archs? CodeGen, Megatron, AI21
\subsection{Background}

Transformer-based deep neural networks have been the cornerstone of advances across domains excelling in language-related tasks~\citep{vaswani_attention_2017,chen_evaluating_2021}. Development and use of Large Language Models (\acp{LLM}) is a substantial evolution in this landscape.
LLMs process inputs formatted as tokens\footnote{tokens are common character sequences that have unique identification through byte pair encoding~\citep{gage_new_1994}.}. When an input sequence of tokens known as a \textit{prompt} is provided, the \ac{LLM} processes it to generate a probability distribution for the next token across the entire vocabulary. The token with the highest probability is chosen, appended to the input sequence, and re-introduced into the \ac{LLM}. This yields a new token, and the process iterates until a complete sequence of tokens, known as a \textit{completion}, is generated.

In the domain of coding, \acp{LLM} exhibit a unique and innovative approach. In addition to natural-language, LLMs are trained on extensive code bases, either in a specific programming language or a mix of languages. The datasets used for these training tasks can often reach into hundreds of gigabytes. The prompts used for these code-trained \acp{LLM} can take various forms, including instructions, comments, code snippets, or a combination of these. Interestingly an \ac{LLM} trained on a corpus of mixed languages often manages to infer the target language from the prompt.

Despite the remarkable abilities of \acp{LLM}, training them from scratch is resource-intensive, requiring massive datasets and numerous parameters. However, there is a practical alternative of fine-tuning pre-trained \acp{LLM} on specialized datasets to cater to specific tasks. Fine-tuning is significantly efficient as it requires a limited number of training epochs. Several LLMs pre-trained for natural language and code either make the weights available, like NVIDIA's MegatronLM~\citep{shoeybi_megatron-lm_2020} or Salesforce's CodeGen models~\citep{nijkamp_conversational_2022}, or provide fine-tuning through an API, like AI21studio's Jurassic-1 (J1) models.\footnote{\url{{https://studio.ai21.com/docs/jurassic1-language-models/\#general-purpose-models}}}~\citep{ai21_jurassic-1_2021}. However, fine-tuning through APIs may come with costs and lacks transparency as the models' weights and parameters are not disclosed.

\rt{Recently, the \ac{LLM} landscape has seen remarkable advancements in terms of larger and more complex models such as GPT-3.5-turbo~\citep{ye2023comprehensive}, GPT-4~\citep{openai_gpt-4_2023}, PALM2~\citep{chowdhery_palm_2022}. These LLMs have not only surpassed their predecessors in their capabilities but have achieved remarkable human task alignment. They have been trained on vast datasets, often comprising gazillions of tokens, pushing the frontier of language-related tasks like code generation, debugging and interpretation. These LLMs have achieved state-of-the-art results, showcasing the potential of machine learning in understanding and emulating human-like text generation.

However, these advances have brought new challenges. Larger models, unlike their open-source counterparts, come with usage fees. %due to their commercial nature. 
Furthermore, these commercial off-the-shelf products LLMs do not guarantee reliability or continuous presence in the public domain. For instance, Code-Davinci-002, a commercial \ac{LLM}, is no longer accessible, highlighting risks of dependency on such models.
Moreover, interfacing with these \acp{LLM} via APIs presents security vulnerabilities, such as the risk of injection attacks, where malicious commands could be inserted and executed. Furthermore, API usage may introduce additional latency, impacting the performance of applications that rely on these LLMs for real-time processing. We refer to these models that are supposedly hundreds of billions of parameters as \textit{large} LLMs.}

%detailed in this prior work,
%In this work, we fine tune pre-trained LLMs to support hardware design in Verilog.
%and investigate the capabilities of such models for supporting hardware design in Verilog. 

\subsection{Prior Work}
Programming is challenging, given the need for human designers to interpret and transform natural language specifications into programming structures.
This motivates the use of \ac{NLP} to transform language to code~\citep{mihalcea_nlp_2006}. Hardware design using Verilog HDL is similar to programming. Prior work explored \ac{NLP} techniques for formal system modeling~\citep{drechsler_generating_2012} and
generating assertions~\citep{harris_glast_2016}, albeit on a small scale. Pearce et al. trained \texttt{DAVE}, a small \ac{LLM} to produce Verilog snippets from template-based natural language descriptions for a limited set of functions~\citep{pearce_dave_2020}. 
GitHub's Copilot was evaluated for security bugs produced during out-of-the-box Verilog completions~\citep{pearce_asleep_2022} and was found to be lacking. 
% This study is a large-scale exploration of the capabilities of \acp{LLM} across more design tasks using an automated evaluation framework.
Our study scales this exploration by assessing LLMs across various design tasks using an automated evaluation framework. Despite the lack of open datasets for Verilog-focused LLM training, the promising outcomes from applying LLMs in programming, and their further potential shown in multi-language code repositories like CodeSearchNet~\citep{codesearchnet}, and large-scale LLMs such as GPT-3.5 and GPT-4, emphasize the possibilities for hardware design languages.
%for automated evaluation. 
%To the best of our knowledge, 
There is no open dataset to train and evaluate LLMs on writing Verilog.

\section{\ac{LLM} Training}
\label{sec:data}
% \todoblock{explain the flow for curating and cleaning the data}
%
%This section describes the steps taken to prepare the Verilog training corpus used to fine-tune LLMs.
% In this section, w
We describe our method for training (or fine-tuning) \ac{LLM} models. % for Verilog code generation.
We begin by describing our curated Verilog datasets, followed by the \ac{LLM} architectures and fine-tuning method.
% Methodology = a *system* of methods, or the study of research methods
%and finally how we evaluate the fine-tuned \ac{LLM} models. 

\subsection{Verilog Training Corpus}

% How you gathered the data

% how you changed the data

% Deduplication
% Filtering (keyword, length)
% Tokenization

% Amount of the data before, and the amount of the data after (or after each step)

Our primary Verilog training corpus comes from open-source Verilog code in public GitHub repositories. Additionally, we also created a dataset of text from Verilog textbooks to understand whether that further improved \ac{LLM} performance.

\paragraph{GitHub Corpus}  
We use Google BigQuery to gather Verilog repositories from GitHub, where it has a snapshot of over 2.8 million repositories. We use a query that looks for keywords such as ``Verilog" and files with  `.v' extension.
%, we downloaded  about 200K files.
%with file size ~3GB. 
We de-duplicated files (using MinHash and Jaccard similarity metrics~\citep{yan_privmin_2017}) 
%with a Jaccard threshold of 0.85 
and filtered files by keeping `.v' files that contain at least one pair of \texttt{module} and \texttt{endmodule} statements. This is done to avoid retaining any .v file without any Verilog. 
The resultant number of .v files are about 50k with a size of ~1GB. 
Finally, we filtered large files (number of characters $\geq$ 20K). This helps in narrowing down the window to which to pay attention to, and hence result in better generalization capability. 
The training corpus from GitHub yielded $\sim$50K files / $\sim$300~MB. %left with a size of 153MB.

\paragraph{Verilog Books Corpus} We downloaded 70 Verilog-based textbooks from an online e-library in PDF format, then extracted text using the Python-based tool pymuPDF which uses optical character recognition to extract text. Depending on the quality of the PDF, the text quality varies. We encountered challenges including improper formatting of lines, unrecognized logical operators, and header/footer floating in-between text. 
We cleaned the text by filtering irrelevant passages (e.g., index, preface, and acknowledgments) and used regular expressions such as: {\footnotesize\verb/module(.*\n*\s*\t*)(\()((?!module)(?!endmodule).*\W*)*endmodule/} 
to identify blocks of prose and associated Verilog snippets. From these blocks, overlapping sliding windows were used to produce training examples.
% \todo{Modified}
%and filters to ensure pair of module and endmodule are present, if a line of code is broken into multiple, we look for immediate trailing semicolons and merge the fragments. 
The final Verilog corpus of textbook-extracted and GitHub code had a size of 400~MB.

\subsection{Baseline LLM Architectures}
Table \ref{tbl:llm-architecture} shows the \acp{LLM} we used and summarizes design parameters, including the number of layers, heads, embedding size (head dimension), context length, and the data source (natural language (NL) and/or code).
%depending on whether they are language-based or code-based models. 
As code-davinci-002 is derived from GPT-3~\citep{chen_evaluating_2021}, it has the same architecture. Since its parameters are not known, so we leave these as \textit{NA}. % in the table.
%175B parameter LLM.

% \begin{table}[h]
% \resizebox{\linewidth}{!}{
%     \begin{tabular}{llccccc}
%         \toprule
%         \multirow{2}{*}{Model} &  \multirow{2}{*}{Dataset}  &\multirow{2}{*}{Parameters} & \multirow{2}{*}{Layers} & \multirow{2}{*}{Heads} & \multirow{2}{4em}{Embedding size} & \multirow{2}{4em}{Context length} \\
%         &&&&&&\\
%         \midrule
%         MegatronLM \citep{} & Natural Language \citep{devlin_bert_2018, trinh_simple_2019, zellers_defending_2020, radford_language_2019} & 355M & 24 & 16 & 64 & 1024 \\
%         \midrule
%         J1-Large & Natural Language \citep{brown_language_2020} & 7B & 32 & 32 & 128 & 4096 \\
%         \midrule
%         \multirow{3}{*}{CodeGen} & \multirow{3}{10em}{Natural Language \citep{gao_pile_2020} + Code \citep{noauthor_bigquery_nodate} } & 2B & 32 & 32 & 80 & 2048 \\
%          &  & 6B & 33 & 16 & 256 & 2048 \\
%          &  & 16B & 34 & 24 & 256 & 2048 \\
%          \midrule
%         \multirow{2}{*}{Code-davinci-002}& \multirow{2}{10em}{Natural Language \citep{brown_language_2020} + GitHub code} & \multirow{2}{*}{175B} & \multirow{2}{*}{96} & \multirow{2}{*}{96} & \multirow{2}{*}{128} & \multirow{2}{*}{12288} \\
%         &&&&&&\\
%         \bottomrule
%     \end{tabular}
% }
% \vspace{0.5em}
% \caption{LLM architecture}
% \label{tbl:llm-architecture}
% \end{table}

\begin{table}[!t]
\caption{Baseline LLM architectures used in our study.}
\tiny
\resizebox{\linewidth}{!}{
    \begin{tabular}{L{3cm}C{0.8cm}C{0.8cm}C{0.8cm}C{1cm}}
    
        \toprule
        % \multirow{2}{*}{Model-Parameters / Pre-Training Data} & \multicolumn{4}{c}{Hyperparameters}\\
        % \cmidrule(lr){2-5}
        % & \multirow{2}{*}{Layers} & \multirow{2}{*}{Heads} & \multirow{2}{4em}{Embedding size} & \multirow{2}{4em}{Context length} \\
        Model-Parameters / Pre-Training Data  & Layers & Heads & Embed. & Context Length\\
        \midrule
        MegatronLM-355M~\citep{shoeybi_megatron-lm_2020} / NL~\citep{devlin_bert_2019, radford_language_2019} & 24 & 16 & 64 & 1024 \\
        \midrule
        J1-Large-7B$^1$ %~\citep{ai21_jurassic-1_2021} 
        / NL~\citep{brown_language_2020} & 32 & 32 & 128 & 4096 \\
        \midrule
        CodeGen-2B~\citep{nijkamp_conversational_2022} / NL~\citep{gao_pile_2020}, Code
        %~\citep{noauthor_bigquery_nodate}
        & 32 & 32 & 80 & 2048 \\
        CodeGen-6B~ /  NL~\citep{gao_pile_2020}, Code & 33 & 16 & 256 & 2048\\
        CodeGen-16B / NL~\citep{gao_pile_2020}, Code & 34 & 24 & 256 & 2048\\
         \midrule
        code-davinci-002~\citep{chen_evaluating_2021}~\label{davinci} /  NL~\citep{brown_language_2020}, Code & NA & NA & NA &  8001\\
         \midrule
        \rt{GPT-3.5-turbo~\citep{ye2023comprehensive}} /  NL~\citep{brown_language_2020}, Code & NA & NA & NA &  4096\\
        \rt{GPT4~\citep{openai_gpt-4_2023}} /  NL, Code & NA & NA & NA &  8000\\
         \midrule
        \rt{PALM2~\citep{chowdhery_palm_2022}} /  NL, Code & NA & NA & NA &  8000\\
        %  \midrule
        % \rt{claude~\citep{anthropic}} /  NL~\citep{brown_language_2020}, Code & NA & NA & NA &  NA\\
        %  \midrule
        % \rt{starcoder~\citep{li_starcoder_2023}} /  NL~\citep{brown_language_2020}, Code & NA & NA & NA &  8192\\
        \bottomrule
    \end{tabular}
    }
% \vspace{0.5em}
\label{tbl:llm-architecture}
\end{table}
%\section{LLM training}
% \todoblock{complete this part, write the description of the systems figure}
\label{sec:training}

\subsection{LLM fine-tuning}
%We consider five different models for fine-tuning as described in \autoref{tbl:llm-architecture}.
We fine-tune five LLMs from~\autoref{tbl:llm-architecture} on our Verilog training datasets. Training the CodeGen LLMs was challenging due to the large number of parameters. At 16-bit precision, the parameters of CodeGen-16B alone require around 30 GB of GPU memory. Further, the fine-tuning process requires even more GPU memory to store the intermediate computations and optimizer states, totaling around 250GB. This necessitates the use of multiple GPUs for successful operation. For instance, in our case, we relied on 3 A100 GPUs to train this model for just one epoch.
%Therefore, 
%to efficiently use the compute resources, 
We use model and data parallelism and strategies for sharding the optimizer states across GPUs similar to  DeepSpeed\footnote{\url{https://huggingface.co/docs/transformers/main_classes/deepspeed}}~\citep{deepspeed} and ZeRO~\citep{ren_zero-offload_2021}. 

We set the training hyperparameters to their defaults as recommended for the ZeRO-3~\citep{rajbhandari_zero-infinity_2021} optimizer state implementation. 
The CodeGen LLMs (2B, 6B, 16B) are fine-tuned for 1 epoch on an HPC cluster with two RTX8000s, four RTX8000s, and three A100s, and training completes in two, four, and six days, respectively. 
Megatron-LM is fine-tuned for 9 epochs using one RTX8000 for 15 hours using the default configuration~\citep{shoeybi_megatron-lm_2020}. We use the off-the-shelf AI21 studio for fine-tuning J1-Large. 
% (estimated training cost 63). (# epochs 20)

\section{LLM Evaluation Setup}
\label{sec:evaluation}

\rt{To gauge the proficiency of Large Language Models (LLMs) in generating high quality Verilog code, we employ two distinct evaluation harnesses. The first harness, fully transparent in its operation, includes a hand-designed problem set (Set I, in~\autoref{tbl:problem_set-1}), a collection of hand-designed test benches, and a comprehensive end-to-end pipeline to determine if the Verilog code, corresponding to a given prompt, adheres to the criteria of functional correctness.

Our second evaluation harness relies on a broader and more diversified set of problems (Set II, in~\autoref{tbl:problems-set-2}) extracted from HDLBits, a popular digital logic learning platform~\citep{wong_projectabout_2019} that allows users to test their solutions against a built-in test bench suite.}

\subsection{Problem Sets}\label{subsec:problem-sets}
\begin{figure}[h]
    \centering
    \includegraphics[width=0.99\linewidth]{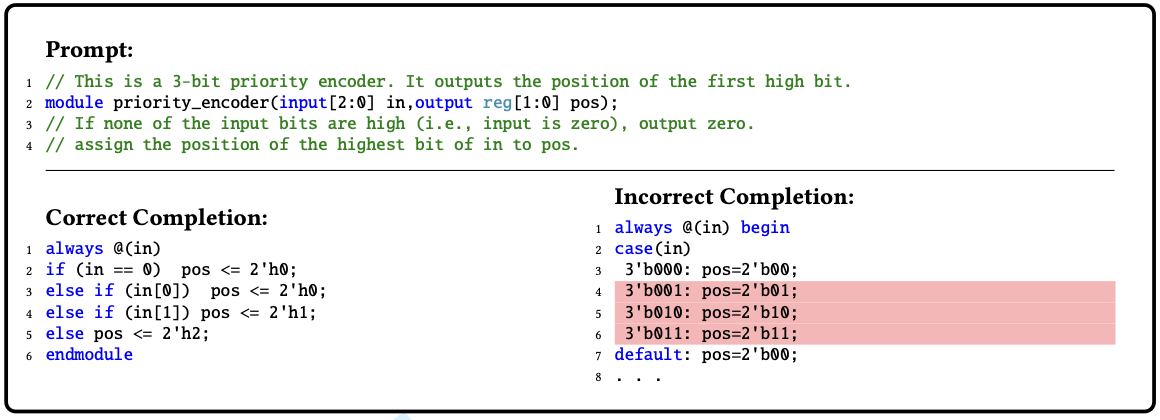}
    \caption{Set I, Basic example - Problem 3: A 3-bit priority encoder. We highlight the \fcolorbox{aired}{aired}{mistake}. Positions are offset by 1.}
\label{fig:basic-example-prompt}
\end{figure}

\begin{figure}[h]
    \centering
    \includegraphics[width=0.99\linewidth]{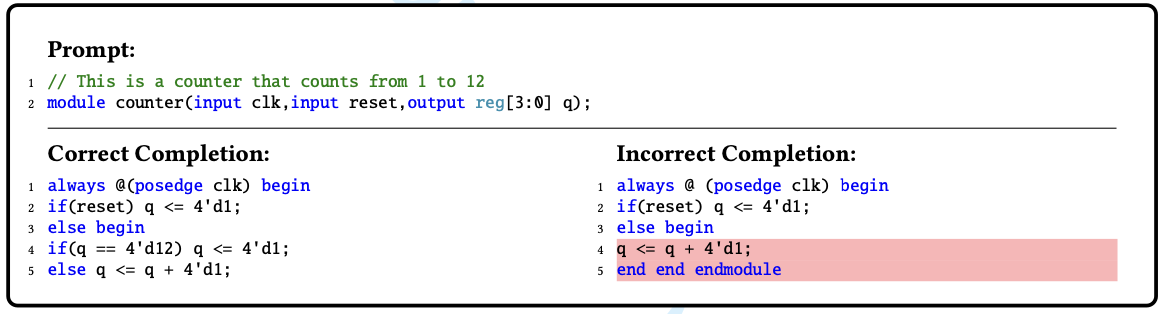}
    
\caption{Set I, Intermediate example - Problem 6: A 1 to 12 counter. We highlight the \fcolorbox{aired}{aired}{mistake}. Counter does not stop at 12.}
\label{fig:comp-counter}
\end{figure}

\begin{figure}[h]
    \centering
    \includegraphics[width=0.99\linewidth]{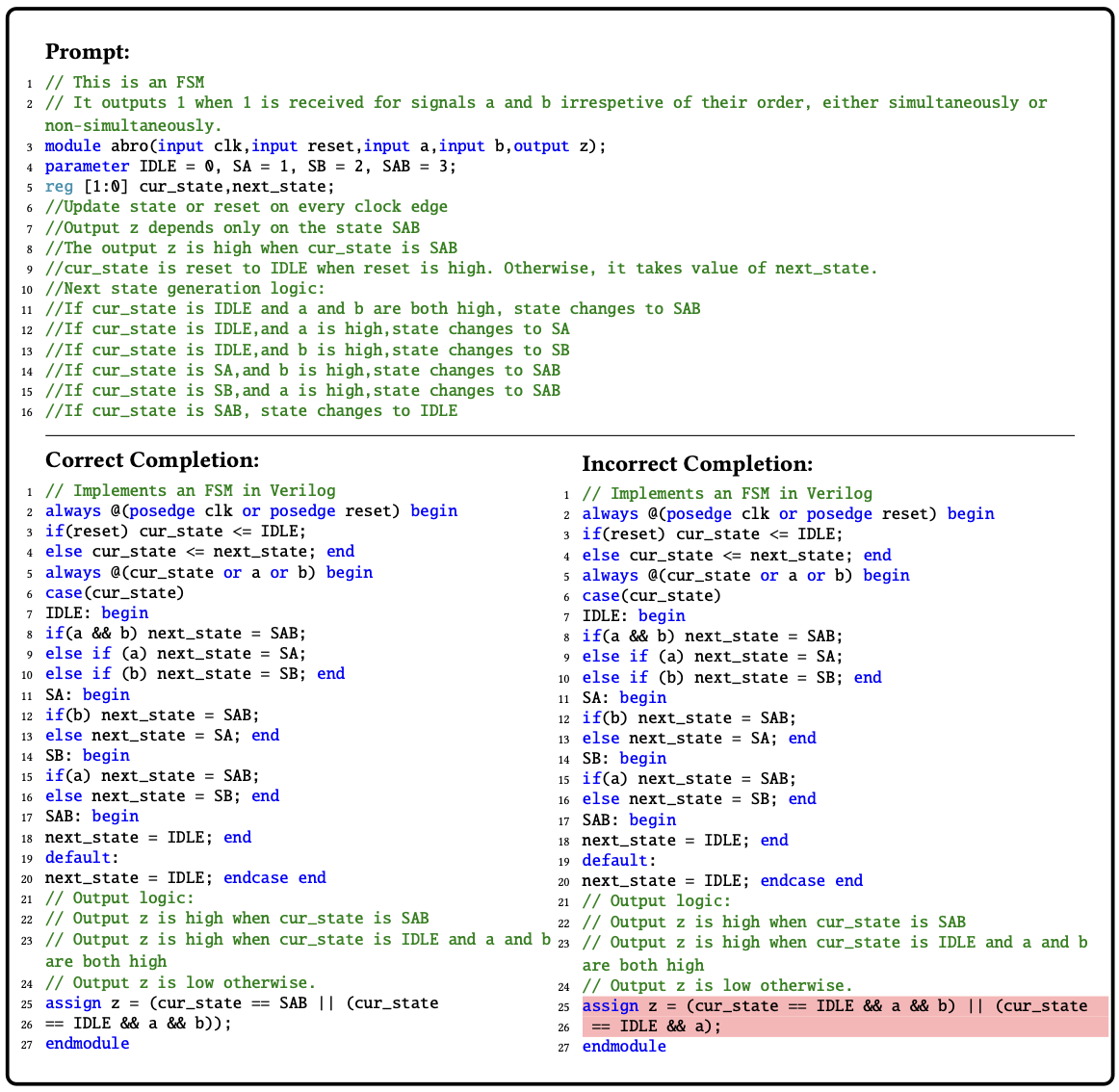}
    
\caption{Set I, We highlight the \fcolorbox{aired}{aired}{mistake}. Output is not assigned to state SAB.}
\label{fig:advanced-example-prompt}
\end{figure}

\noindent \textbf{Problem Set I} is composed of 17 unique Verilog challenges, which have their roots in classroom exercises and examples from the HDLBits platform. Each problem has an assigned difficulty level, which is depicted in~\autoref{tbl:problem_set-1}. These challenges cover a wide range of design concepts, including combinational and sequential logic designs, finite state machines with varying requirements, operations such as permutation, shift left, and rotate, and basic components like a multiplexer (MUX), Random Access Memory (RAM), Linear Feedback Shift Register (LFSR), adders, and counters. Examples of basic, intermediate, and advanced problems are presented in Fig.~\ref{fig:basic-example-prompt}-\ref{fig:advanced-example-prompt} respectively. These have been obtained using CodeGen-16B-FT and have been edited for clarity and brevity.\\

\rt{\noindent \textbf{Problem Set II} significantly expands on this initial set by integrating an extensive range of problems from HDLBits~\citep{wong_projectabout_2019}. This expansion enriches our dataset to encompass a total of 181 problems. The enlarged dataset introduces new challenges distributed across four difficulty levels and multiple unique categories, each dedicated to a specific facet of hardware design and Verilog syntax. The dataset spans difficulty levels ranging from \textit{Getting Started} to \textit{Verilog Language} to \textit{Circuits} and \textit{Verify Bugs}, shown in~\autoref{tbl:problems-set-2}. Within these levels, categories extend from basic principles such as gates to more advanced concepts like finite state machines and cellular automata, thus presenting a comprehensive view of hardware design and Verilog syntax. }\\
\newsavebox{\mintedboxtwo}
\newsavebox{\mintedboxone}

\definecolor{blond}{rgb}{0.98, 0.94, 0.75}
\definecolor{lightergray}{RGB}{220, 220, 220}

\begin{figure}[h]
    \centering
    \includegraphics[width=0.99\linewidth]{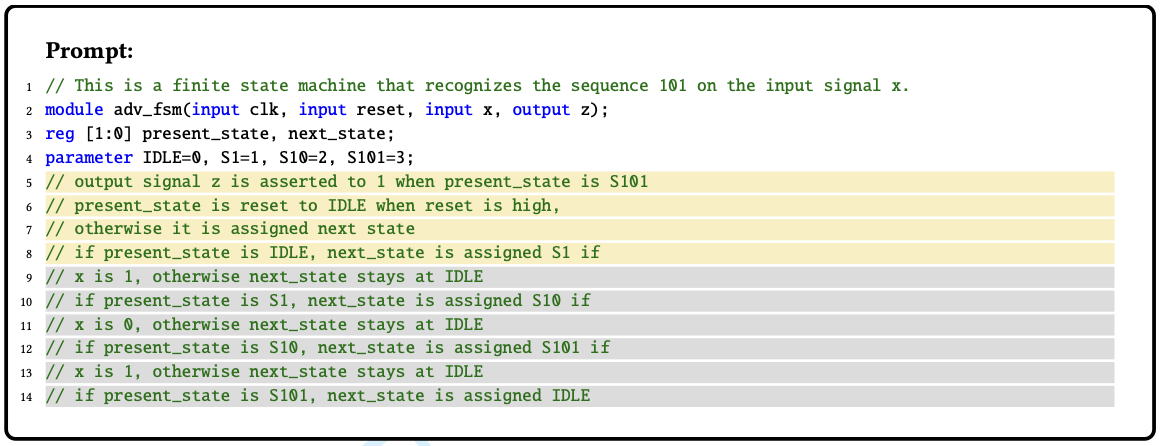}
    \caption{Varying the prompt details: Low, Medium and High. Set I, Problem 15.}
\label{fig:prompt-details}
\end{figure}

% \begin{figure}[h]
% \begin{AIbox}{}
% \textbf{Prompt:}
% \begin{lstlisting}[language=Verilog,, linebackgroundcolor={\ifnum\value{lstnumber}>4
%                 \ifnum\value{lstnumber}<9
%                     \color{blond}
%                 \fi
%             \fi
%             \ifnum\value{lstnumber}>8
%                 \ifnum\value{lstnumber}<16
%                     \color{lightergray}
%                 \fi
%             \fi} ]
% // This is a finite state machine that recognizes the sequence 101 on the input signal x. 
% module adv_fsm(input clk, input reset, input x, output z); 
% reg [1:0] present_state, next_state;
% parameter IDLE=0, S1=1, S10=2, S101=3;
% // output signal z is asserted to 1 when present_state is S101
% // present_state is reset to IDLE when reset is high, 
% // otherwise it is assigned next state
% // if present_state is IDLE, next_state is assigned S1 if 
% // x is 1, otherwise next_state stays at IDLE
% // if present_state is S1, next_state is assigned S10 if
% // x is 0, otherwise next_state stays at IDLE 
% // if present_state is S10, next_state is assigned S101 if 
% // x is 1, otherwise next_state stays at IDLE 
% // if present_state is S101, next_state is assigned IDLE
% \end{lstlisting}
% \end{AIbox}
% \caption{Varying the prompt details: Low, Medium and High. Set I, Problem 15.}
% \label{fig:prompt-details}
% \end{figure}

\subsection{LLM Inference}

The LLM input is a prompt from the problem set in \autoref{subsec:problem-sets}. 
We truncated completed code at keywords \texttt{end} and \texttt{endmodule} and pass the solution to the appropriate evaluation harness for checking its compilation and functional correctness.

\subsubsection{Input Parameters}

Each query to the \ac{LLM} includes a prompt, a sampling temperature ($t$), and a number of completions per prompt ($n$). \\

\noindent \textbf{Prompts:} For the hand-designed Problem Set I, each problem is accompanied by three prompts of increasing detail: low (L), medium (M), and high (H). 
% We provide three prompts with increasing detail, i.e., low (L), medium (M), and high (H). % are provided.
Prompt L has an initial comment describing the function of the module and the module header with name and inputs/outputs with types. 
We declare internal signals. % are also declared. 
M includes L plus comments that describe the function using signal names. 
H replaces and/or appends comments in M with more detail and resembles pseudo-code instead of a predominantly natural language specification. \autoref{fig:prompt-details} is an example for Problem 15. 
L has no lines highlighted (the prompt is lines 1--4). 
M includes L and lines highlighted yellow (the prompt is lines 1--8). 
H includes M and lines in gray (the prompt is lines 1--15).

\begin{figure}[h]
    \centering
    \includegraphics[width=0.99\linewidth]{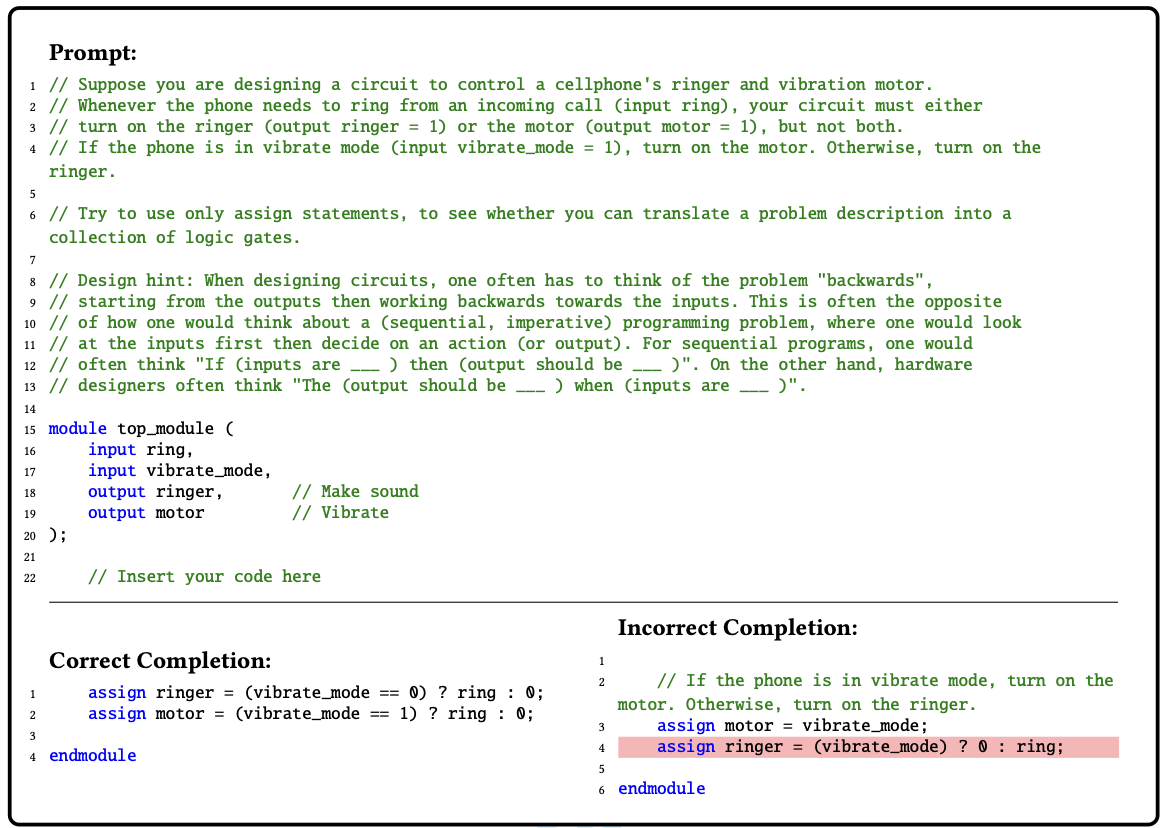}
    \caption{Set II, Vibrate \& ring problem. Difficulty: Circuits (Combinational), Basic category. We highlight the \fcolorbox{aired}{aired}{mistake}. \texttt{motor} turns on even with \texttt{ringer} set to 0 and \texttt{vibrate\_mode} set to 1}
\label{fig:ringer}
\end{figure}

\rt{For Problem Set II, we formulate the prompts in alignment with the problem descriptions provided on the platform. A typical prompt for each problem starts with a set of comments, abstracted from the high-level problem description available on the platform. Following this is the module header, tagged as \texttt{top\_module}, which includes inputs and outputs, each with its defined type. Finally, a comment line prompts the insertion of the problem-solving code.

An illustrative example of this format is depicted in~\autoref{fig:ringer}, applied to the \textit{vibrate\&ring} problem. This prompt features a detailed level of description. The lines 1-13 contain the initial comments, which describe the problem function. This section uses signal names and high-level specifications. The following segment (lines 15-20) includes the module header skeleton, which comprises the module's name and input/output data. And line 22 holds a single line comment instructing the user to insert their code.}

\noindent \textbf{Sampling temperature $(t)$:} A higher value means that the LLM takes more risks and yields more creative completions. We use $t \in \{0.1,0.3.0.5,0.7,1\}$.
    
\noindent \textbf{Completions per prompt $(n)$:} \rt{For each prompt, LLM generates $n$ completions where $n \in \{1,10,25\}$. For J1-Large and PALM2, we skip $n=25$ because they do not support this value. For GPT4, we only use $n=1$ due to the high cost of usage.}
%\end{itemize}

\noindent \textbf{max\_tokens:} \rt{The maximum number of tokens generated for each completion was set to 300 for all autocomplete language models (Megatron-LM, CodeGen, Code-davinci-002), except for J1-Large. J1-Large had a token limit of 256. We used the default nucleus sampling probability mass (\texttt{top\_p}) setting. For larger models like GPT3.5, PALM2, and Claude, which generate responses in a conversational style, we increased the maximum tokens to 900 per completion, as these models typically consume more tokens compared to autocomplete ones. However, for GPT4, the web interface doesn't provide an option to adjust the token limit.}

\subsection{Test benches}
For Problem Set I, we developed a test bench to validate the functional correctness of each solution. The test benches exercise the designs for corner cases and are exhaustive for basic and some intermediate cases. 
For the remaining cases,
%and similar to prior work %\citep{chen_evaluating_2021},
the test benches are analogous to unit tests. 
This keeps the evaluation time reasonable, e.g., for the RAM module, the data width is 8 and the address width is 6 in the prompt; an exhaustive test bench requires $2^{14}$ test inputs. % and would take longer to simulate. 
In some cases, specifications in the prompts are ambiguous and thus can yield several correct responses. For example, 
% If the specifications in the prompts are not detailed enough, it enables a variety of correct responses. 
% For example, 
when one does not specify whether a reset should be synchronous or asynchronous. % in an FSM.
We compile and simulate completed Verilog with Icarus Verilog v11.0~\citep{williams_icarus_2023}.

\rt{For Problem Set II, we turn to the HDLBits online judge system. The online system features an "upload and compile" feature that accepts the completed Verilog as a solution, synthesizes the Verilog code using Quartus, and verifies its correctness through simulations run by Modelsim. At the end of the execution cycle, the system returns a detailed summary of the execution. Only if the completed Verilog passes all the tests within the test-bench for the problem will it return a status of "Success!". In other cases, responses include "Compile Error", "Simulation Error", and "Incorrect", each indicating different stages of failure in the evaluation process.

Despite limited visibility into the specifics of the test benches used for the problems, the feedback from the online system allows us to assess the Large Language Models' proficiency in producing functionally accurate and high-quality Verilog code. }

\begin{table}[t]
\caption{Problem Set I}
% \footnotesize
\centering
\begin{tabular}{l l l} 
 \hline
 Prob. \# & Difficulty & Description \\ 
 \hline
 1 & Basic & A simple wire \\ 
 2 & Basic & A 2-input and gate \\
 3 & Basic &  A 3-bit priority encoder \\
 4 & Basic & A 2-input multiplexer \\
 5 & Intermediate & A half adder  \\
 6 & Intermediate & A 1-to-12 counter  \\
 7 & Intermediate & LFSR with taps at 3 and 5 \\
 8 & Intermediate & FSM with two states \\
 9 & Intermediate & Shift left and rotate \\
 10& Intermediate & Random Access Memory\\
 11& Intermediate & Permutation  \\
 12& Intermediate & Truth table \\
 13& Advanced & Signed 8-bit adder with overflow \\
 14& Advanced & Counter with enable signal \\
 15& Advanced & FSM to recognize `101' \\
 16& Advanced & 64-bit arithmetic shift register \\
 17& Advanced & ABRO FSM$^{*}$ %\citep{potop-butucaru_compiling_2007}
 \\
 \hline
\end{tabular}
\\ $^{*}$\textit{from Potop-Butucaru, Edwards, and Berry's ``Compiling Esterel''}

% \vspace{-1em}

\label{tbl:problem_set-1}
\end{table}

\begin{table*}
\centering 
\caption{Problem Set II - Extended set of 164 problems derived from HDLBits~\citep{wong_projectabout_2019}}
\begin{tabularx}{\textwidth}{|m{0.1\textwidth}|m{0.1\textwidth}|m{0.04\textwidth}|X|}
\hline
Difficulty & Category & Count & Problem Description \\ \hline
Getting Started & Getting Started & 2 & Getting Started, Output Zero \\ \hline
\multirow{5}{*}{Verilog} & Basics & 8 & Simple/Four wires, Inverter, AND, NOR, XNOR, Declare wires, 7458 chip \\ \cline{2-4}
 & Vectors & 9 & Vectors, Vectors (  detail), Vector part select, Bitwise operators, Four-input gates, Vector concatenate, Vector reversal 1, Replicate, More replication \\ \cline{2-4}
 & Module Hierarchy & 9 & Modules, Connect ports by position, Connect ports by name, Three modules, Modules and vectors, Adder 1, Adder 2, Carry-select, Adder-subtractor \\ \cline{2-4}
 & Procedures & 8 & Always blocks (combinational), Always blocks (clocked), If statement, If statement latches, Case statement, Priority encoder, Priority encoder with casez, Avoiding latches \\ \cline{2-4}
 & More Features & 7 & Conditional ternary, Reduction operators, Reduction: Wider gates, Combinational for-loop: Vector reversal 2, Combination for-loop: 255-bit count, Generate for-loop: 100-bit adder 2, Generate for-loop: 100-digit BCD adder \\ \hline
\multirow{4}{*}{Circuits} & Basic & 17 & Wire, GND, NOR, Another, Two gates, More gates, 7420 chip, Truth tables, Two-bit equality, Simple circuits A, B, Combine circuits A, B, Ring or vibrate?, Thermostat, 3-bit count, Gates and vectors, longer vectors \\ \cline{2-4}
(Combination) & Multiplexers & 5 & 2-to-1 mux, 2-to-1 bus mux, 9-to-1 mux, 256-to-1 mux, 256-to-1 4-bit mux\\ \cline{2-4}
 & Arithmetic Circuits & 7 & Half adder, Full adder, 3-bit adder, Signed addition overflow, 100-bit binary adder, 4-digit BCD adder \\ \cline{2-4}
 & K-Map to Circuit & 8 & 3/4-variable, Minimum SOP and POS, K-map, K-map  with a mux \\ \hline
\multirow{8}{*}{Circuits } & Latches and Flip-Flops & 18 & DFFs, DFF (reset), DFF (reset value), DFF (asynchronous), DFF (byte enable), D Latch, DFF,  DFF+gate, Mux and DFF, DFFs and gates, Circuit from truth table, Detect edge/both edges, Edge capture register, Dual-edge triggered FF \\ \cline{2-4}
& Counters & 8 & Four-bit binary counter, Decade counter, Decade counter again, Slow decade counter, Counter 1-12, Counter 1000, 4-digit decimal counter, 12-hour clock \\ \cline{2-4}
 (Sequential)  & Shift Registers & 9 & 4-bit shift register, Left/right rotator, Left/right arithmetic shift by 1 or 8, 5-bit LFSR, 3-bit LFSR, 32-bit LFSR, Shift register, Shift register, 3-input LUT \\ \cline{2-4}
 & Cellular Automata & 3 & Rule 90, Rule 110, Conways Game of Life 16x16 \\ \cline{2-4}
 & FSM & 33 & FSM 1 (asynchronous), FSM 1 (synchronous), FSM 2 (asynchronous), FSM 2 (synchronous), Simple state transitions 3, Simple one-hot state transition 3, FSM 3 (asynchronous), FSM 3 (synchronous), Moore FSM, Lemmings 1, Lemmings 2, Lemmings 3, Lemmings 4, One-hot FSM, PS/2 packet parser, PS/2 packet parser and datapath, Serial receiver, Serial receiver and datapath, Serial receiver with parity check, Sequence recognition, Q8: Design Mealy FSM, Q5a: Serial twos complementer (Moore FSM), Q5b: Serial twos complementer (Mealy FSM), Q2a, Q2b, Q3a, Q3b: FSM, Q3c: FSM logic, Q6b: FSM next-state logic, Q6c: FSM one-hot next-state logic, Q6: FSM, Q2a: FSM, Q2b: One-hot FSM equations  \\  \cline{2-4} 
 & Larger Circuits & 7 & Counter with period 1000, 4-bit shift register and down counter, FSM: Sequence 1101 recognizer, FSM: Enable shift register, FSM: The complete FSM, The complete timer, FSM: One-hot logic equations \\ \hline
Verify Bugs & Read Simulations \& Find bugs & 5 & Mux2, NAND, Mux4, Add/subtract, Case statement \\ \hline
\end{tabularx}
\label{tbl:problems-set-2}
\end{table*}

\section{LLM Evaluation and Results}
% \todoblock{Working on it now ..}
\label{sec:Results}
% We now discuss the evaluation metrics and the research questions answered with the results of this study.
\subsection{Research Questions}
% This section introduces experiments, metrics,  and results. %The experiments will evaluate the accuracy and efficiency of the fine-tuned LLMs for Verilog generation. 
We examine the quality of generated Verilog code for scenarios and test benches from~\autoref{subsec:problem-sets}. This analysis seeks to address the following research questions (RQs): 
\begin{itemize}
\item 
\textbf{RQ1}. How well do `base' LLMs perform on the Verilog generation set?
\item 
\textbf{RQ2}. Does fine-tuning LLMs improve that performance?
\item 
\textbf{RQ3}. Are larger LLMs with more parameters better?
\item 
\textbf{RQ4}. Does variability in problem description impact quality and the number of correct completions?

\rt{\item \textbf{RQ5}. How does the performance of fine-tuned model stack up against \textit{large} LLMs such as GPT4, GPT-3.5-turbo, claude and PALM2 in producing Verilog code for problems of different complexities?
\item \textbf{RQ6}. At what levels of problem difficulty do large LLMs excel, and where might they need enhancement to either meet or surpass the best performing model?
 \item \textbf{RQ7}.  Can incorporating diverse sources of training data, like educational resources (such as textbooks), lead to better model performance?
}
\end{itemize}

%in each problem difficulty level associated test bench.

\begin{table}[!t]

\centering
\caption{Pass@(scenario*$n$) at $n$=10 for compiled completions (Pass=Compiling), PT = Pre-trained, FT = Fine-Tuned, Problem Set I. Bold reflects the (best) highest performance for that difficulty.}
\begin{tabular}{lcccc}
\toprule
\textbf{Model} & \textbf{Model Type} & Basic &  Intermediate &  Advanced \\
\midrule
\multirow{2}{*}{MegatronLM-345M} & PT & 0.000 &         0.000 &     0.000 \\
 & FT &  0.730 & 0.391 &     0.165 \\
 \midrule
\multirow{2}{*}{CodeGen-2B}  & PT    &  0.080 &         0.065 &     0.176 \\
   & FT   &  0.902 & 0.612 &     0.592 \\
\midrule
\multirow{2}{*}{CodeGen-6B} & PT      &  0.052 &         0.152 &     0.187 \\
& FT      &  \textbf{0.987} &   0.689 &     \textbf{0.599} \\
\midrule
\multirow{2}{*}{J1-Large-7B} & PT      &  0.182 &         0.176 &     0.108 \\
& FT      &  0.882 &    0.635 &     0.588 \\
\midrule
\multirow{2}{*}{CodeGen-16B} & PT     &  0.132 &         0.203 &     0.240 \\
& FT     &  0.942 &    \textbf{0.728} &     0.596 \\
\midrule
code-davinci-002 & PT   &  0.847 &         0.452 &     0.569 \\
\bottomrule
\end{tabular}
\label{tbl:compiled}
\end{table}

\begin{table*}
\centering
\caption{Pass@(scenario*$n$) at $n$ =10 for test bench passing completions (Pass=Passed Functional Tests), PT = Pre-trained, FT = Fine-Tuned, Problem Set II. Bolded value in each test column reflects the (best) highest performance for that problem set and difficulty.}
% \footnotesize
\begin{tabular}{lm{0.9cm}c|ccc|ccc|ccc}
\toprule
\multirow{3}{*}{\textbf{Model}} & \multirow{3}{1em}{\textbf{Model Type}}  & \multirow{3}{5em}{\textbf{Inference Time (s)}} & \multicolumn{3}{c}{\textbf{Basic}} & \multicolumn{3}{c}{\textbf{Intermediate}} & \multicolumn{3}{c}{\textbf{Advanced}}  \\
\cmidrule(lr){4-12}
        &  & &L & M & H & L & M & H & L & M & H  \\
\midrule
\multirow{2}{*}{MegatronLM-355M} & PT &   3.628    &   0.000 &          0.000 &          0.000 &                 0.000 &                 0.000 &                 0.000 &             0.000 &             0.000 &             0.000\\
 & FT &          0.175 &          0.170 &   0.591   &    0.245 &                 0.043 &                 0.018 &                 0.025 &             0.000 &             0.000 &             0.000 \\
 \midrule
 \multirow{2}{*}{CodeGen-2B} & PT &      1.478  &          0.000 &          0.000 &          0.000 &                 0.000 &                 0.000 &                 0.000 &             0.000 &             0.016 &             0.020\\
 & FT       &          0.665 & 0.835       &  0.350 &          0.630 &                 0.130 &                 0.092 &                 0.163 &             0.132 &             0.048 &             0.068 \\
 \midrule
 \multirow{2}{*}{CodeGen-6B} & PT  &  2.332     &          0.000 &          0.000 &          0.000 &                 0.000 &                 0.000 &                 0.013 &             0.000 &             0.000 &             0.000\\
  & FT       &          0.710 &    \textbf{1.000}    &  0.500 &          0.760 &                 0.135 &                 0.150 &                 0.168 &             \textbf{0.284} &             0.164 &             0.164\\
 \midrule

\multirow{2}{*}{J1-Large-7B} & PT   & 7.146      &          0.044 &          0.058 &          0.067 &                 0.000 &                 0.000 &                 0.021 &             0.000 &             0.000 &             0.000 \\
 & FT         &   2.029    &   0.388 &          0.283 &          0.342 &                 0.125 &                 0.075 &                 0.200 &             0.000 &             0.000 &             0.000 \\
\midrule

 \multirow{2}{*}{CodeGen-16B} & PT      &    2.835    &  0.000 &          0.085 &          0.055 &                 0.035 &                 0.003 &                 0.045 &             0.012 &             0.000 &             0.016\\
 & FT      &    1.994    &  0.745 &          \textbf{0.720} &          0.745 &                 \textbf{0.213} &                 \textbf{0.270} &                 \textbf{0.255} &             0.246 &             \textbf{0.290} &             0.294\\
 \midrule
code-davinci-002 & PT   &   3.885  &     0.520 &          0.685 &          \textbf{0.775} &                 0.175 &                 0.200 &                 0.150 &             0.156 &             0.184 &             \textbf{0.344}\\
\bottomrule
\end{tabular}
\vspace{-0.5em}
\label{tbl:results}
\end{table*}
% Each model is asked to generate N code suggestions for the input problem ac 
% \subsection{Evaluation metric}
\subsection{Evaluation on custom-designed problem set}
In the first study, we measure generated code quality of the fine-tuned models \& their pre-trained versions using problem Set I described in~\autoref{sec:evaluation}. 
A scenario is a combination of problems across difficulties and description levels. 
We query the models with all prompt $\times$ $t$ $\times$ $n$ combinations. 
For fairness, we present each model's ``\textit{best results}'' by focusing on the completions generated with the $t$ for each model for which their completions were most successful at compiling and passing the functional tests (for each problem difficulty and description level). 
% Given a prompt including a problem description, the LLM samples tokens until stop words such as \{\texttt{endmodule}, \texttt{end}\} are found, and return the completions. 
% To evaluate the quality of completions, we sweep over temperature  %$t\in\{0.1,0.3,0.5,0.7\}$ 
% and \# of completions, % $n\in\{1,10,25\}$ 
% while keeping sampling probability, \texttt{top\_p}$=1$, \texttt{max\_token}{=300}. 
% 
% I.e. For each difficulty level and each problem, we find the best-performing aggregate suggestions for each model parameters and present those as the performance value.
% 
We present these \textit{best results} for $n=10$ in \autoref{tbl:compiled} and \autoref{tbl:results}. 
\autoref{tbl:compiled} shows the proportion of completions that compile and \autoref{tbl:results} shows the proportion of completions that pass functional tests, for the completions produced by a given temperature setting that resulted in the most successful completions for each scenario.
As in prior work~\citep{nijkamp_conversational_2022}, we characterize the model performance with the Pass@$k$ metric, where $k$ is the number of problems in a \textit{scenario} times $n$, the number of suggestions per problem. A higher Pass@$k$ indicates a relatively `better' result. 
For compilation (\autoref{tbl:compiled}), the Pass@$k$ metric reflects the proportion of completions that compile.
For functional tests, this metric is the fraction of the $k$ code samples that pass. 

% \subsection{Results}
% We measure the quality of the code generated by LLMs using problem sets described in~\autoref{sec:evaluation}. Given a prompt including a problem description, the LLM samples tokens until stop words such as \{\texttt{endmodule}, \texttt{end}\} are found, and return the completions. To evaluate the quality of completions, we sweep over temperature  %$t\in\{0.1,0.3,0.5,0.7\}$ 
% and \# of completions, % $n\in\{1,10,25\}$ 
% while keeping sampling probability, \texttt{top\_p}$=1$, \texttt{max\_token}{=300}. 

% results by selecting the completions with the highest Pass@(\textit{scenario}*$n$) across temperature and for $n=10$. 
Table~\ref{tbl:results} reports the inference time for each \acp{LLM} query, including communication time with a remote server if required. % For remote models (e.g., J1-Large-7B, code-davinci-002) this includes the round-trip time to communicate to the remote server.
These results are after fine-tuning the model using the training corpus from GitHub only. We discuss the case for fine-tuning on GitHub and PDFs combined as an ablation study in the discussion. Fine-tuned CodeGen-16B LLM outperforms all LLMs. %across the problems. 
%In general 
All fine-tuned LLMs outperform their pre-trained counterparts. %This observations remains even for the model pre-trained on programming languages. 
%The pre-trained code-davinci-002 LLM performs as well as CodeGen-16B. However, it fails as the difficulty level increases. 
%Therefore, based on our study, 
%The Verilog synthesis capability of the base LLMs are of lower quality as that of their fine-tuned counterparts 
[\textbf{Ans. RQ1 and RQ2}].   

\subsubsection*{Completions vs. Temperature ($t$)}%Verilog Synthesis as a function of variation in temperature ($t$)}
%To demonstrate the effect of change in sampling temperature on the model Verilog synthesis capability, 
\autoref{fig:t-n-change-old} summarizes the Pass@(\textit{scenario}*$n$) metric for our experiments sweeping temperature. % for $n=10$ and temperature, $t=\{0.1,0.3,0.5,0.7\}$ and for all LLMs. %To summarize,  
Pass@(\textit{scenario}*10) has the highest value for $t=0.1$ and degrades exponentially with temperature. The LLM  generates accurate solutions at low temperatures
%, and is preferable when less compute is available 
and accurate synthesizable codes are expected from fewer candidates. 

\subsubsection*{Completions vs. \# Completions/Prompt ($n$)}%Impact of variation in $n$ on the synthesis}
We study synthesis quality as a function of completions/prompt. 
The right-hand panel in \autoref{fig:t-n-change-old} shows the Pass@(\textit{scenario}*$n$) for all LLMs. Pass@(\textit{scenario}*$1$) is better than Pass@(\textit{scenario}*$10$). This improves as the number of completions increases. %We hypothesize that 
This is because the number of candidate solutions at low temperatures increases, increasing the completions passing the test benches. $n=10$ is good for all difficulty levels.

\subsubsection*{Completions vs. LLM Size} %Impact of model scale}
\autoref{fig:t-n-change-old} show that LLMs with more parameters (CodeGen-16B, code-davinci-002) outperform LLMs with less parameters such as Megatron-355M and CodeGen-2B. These LLMs yield  more completions that pass test benches and more correct completions. [\textbf{Ans. RQ3}].

\subsubsection*{Completions vs. Prompts} %Verilog synthesis as a function of problem description}
Prompt quality impacts the LLM generation quality. We study the effect of variations in the prompt description at two levels: 
How do the difficulty of the prompt?
And, how does the description of the prompt impact code completions? 

\begin{figure}[t]
    \centering
    \includegraphics[width=0.87\linewidth]{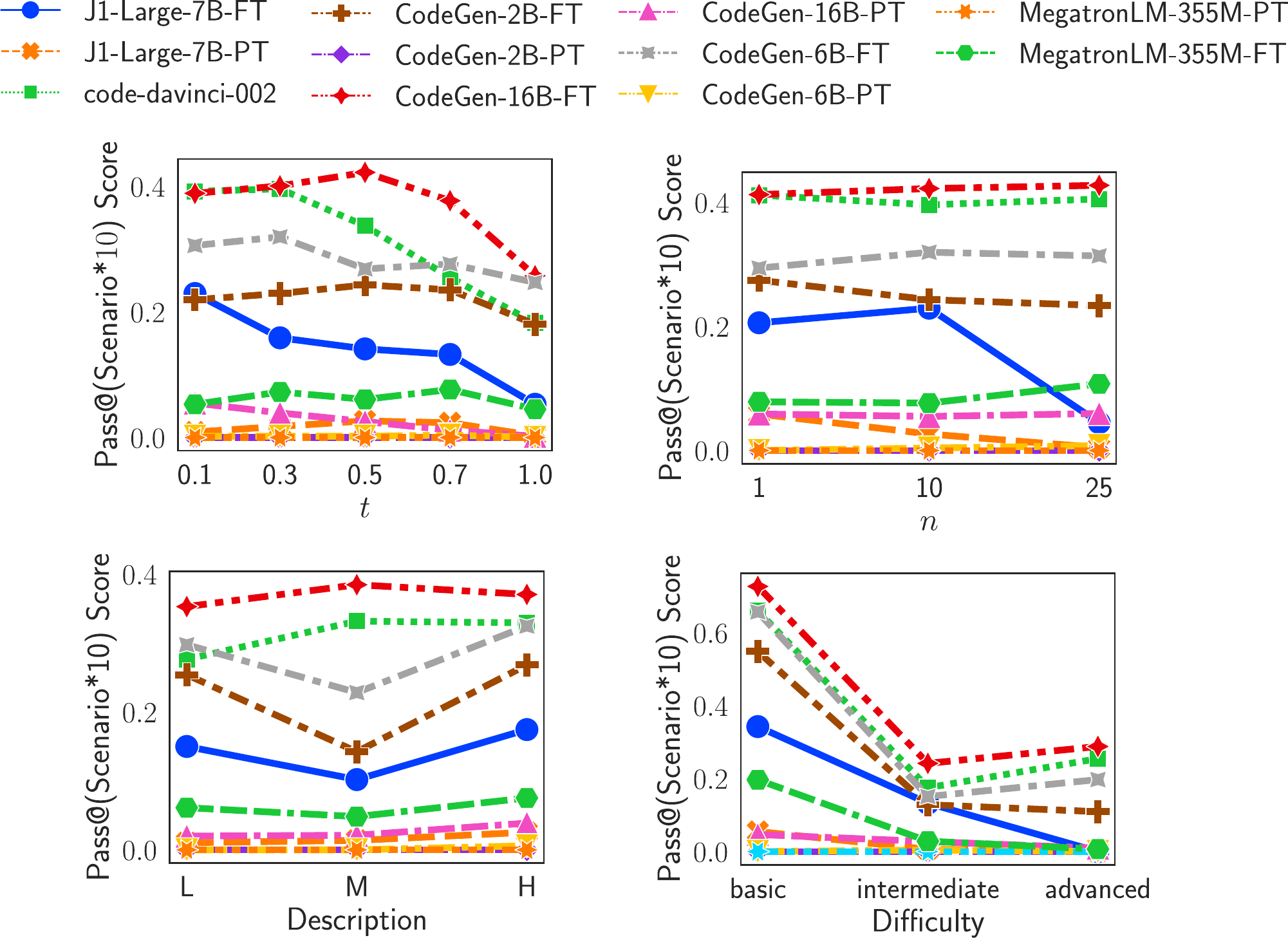}
    % \vspace{-1.5em}
    \caption{Pass@(scenario*$n$) at n=10 for scenarios passing test benches across temperature ($t$), completions per prompt ($n$), prompt description, and problem difficulties for Problem Set I (~\autoref{tbl:problem_set-1}). Comparing the various fine-tuned models alongside their pre-trained version. Higher is better.}
    \label{fig:t-n-change-old}
\end{figure}

We use Pass@(\textit{scenario}*$10$) as the metric.
%With the difficulty level \textit{basic},\textit{intermediate}, and \textit{advanced}, 
The right-hand side panel in the the bottom row in the \autoref{fig:t-n-change-old} shows that the Pass@(\textit{scenario}*$10$) decreases with increasing prompt difficulty.
Simple problems such as AND are easy to translate to Verilog, as opposed to advanced problems such as LFSR. %Similarly, the impact of truncating the problem description. 
The left-hand side panel in \autoref{fig:t-n-change-old} shows that the number of correct solutions decreases with terse prompts. [\textbf{Ans. RQ4}].

% \begin{figure}[t]
% % \setlength\abovecaptionskip{-0.05\baselineskip}
%     \centering
%     \includegraphics[width=0.9\linewidth]{figures/N_temp_var_5.pdf}
%     % \vspace{-1.5em}
%     \caption{Pass@(scenario*$n$) for scenarios passing test benches across temperature ($t$) and completions per prompt ($n$). Higher is better.}
%     \label{fig:t-n-change}
% \end{figure}

% \begin{figure}[t]
% % \setlength\abovecaptionskip{-0.05\baselineskip}
%     \centering
%     \includegraphics[width=0.9\linewidth]{figures/desc_diff_var_1.pdf}
%     % \vspace{-1.5em}
%     \caption{Pass@(scenario*$n$) for scenarios passing test benches across problem difficulties and description levels. Higher is better.}
%     % \vspace{-0.3em}
%     \label{fig:description-difficulty}
% \end{figure}

\subsection{Evaluation on chatGPT and other emerging LLMs}

\begin{review}
% Next we examine the proficiency of instructional LLMs in generating Verilog code. This analysis seeks to address the following research questions (RQs):

\begin{figure}[t]
    \centering
    \includegraphics[width=0.87\linewidth]{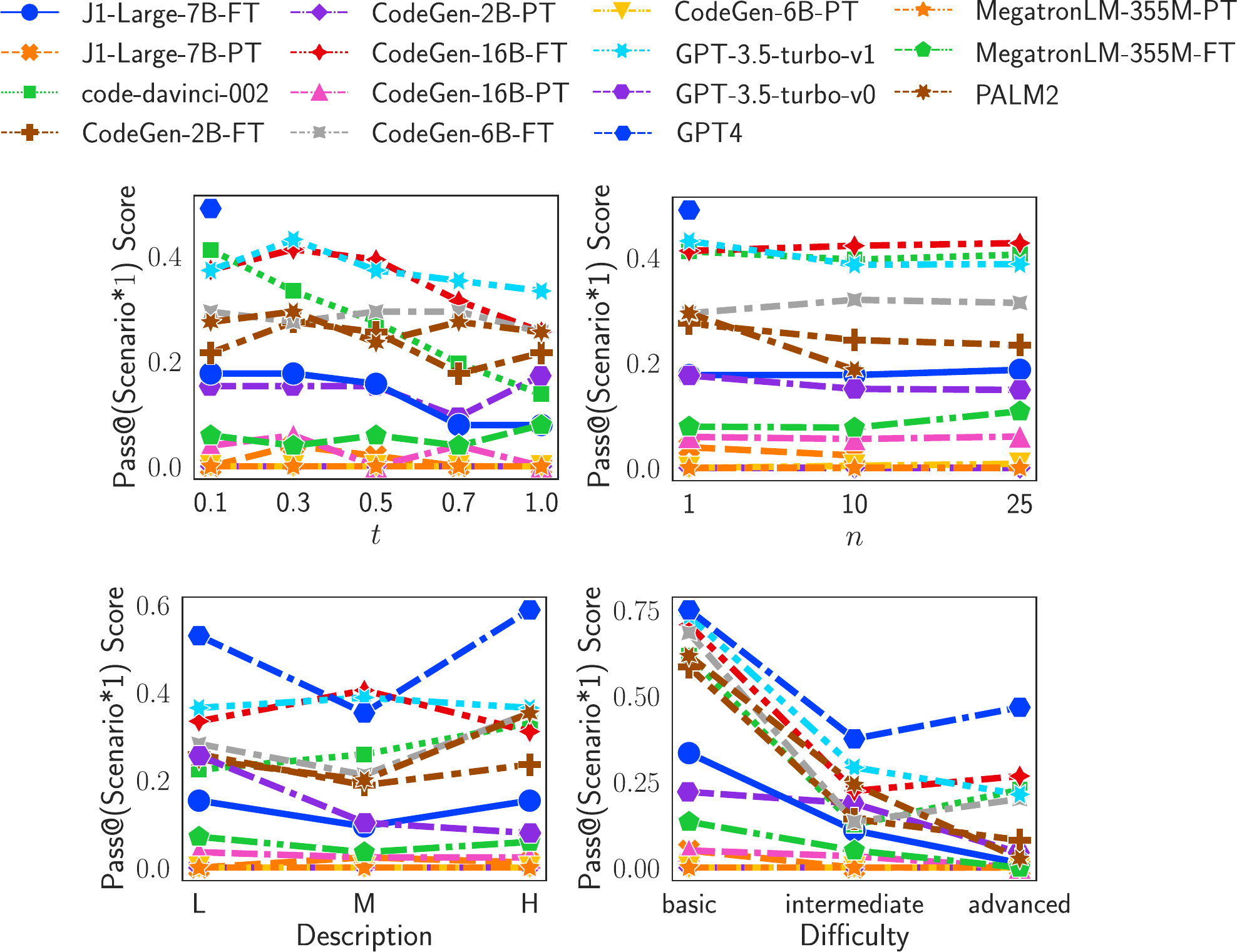}
    % \vspace{-1.5em}
    \caption{Pass@(scenario*$n$) at n=1 for scenarios passing test benches across temperature ($t$), completions per prompt ($n$), prompt description, and problem difficulties for problem Set I (~\autoref{tbl:problem_set-1}). Comparing GPT-3.5-turbo, PALM2, and GPT4 against fine-tuned models. Higher score is better.}
    \label{fig:t-n-change}
\end{figure}
% \begin{itemize}

% \item RQ5: How does the performance of the refined CodeGen-16B-FT model stack up against instructional LLMs such as GPT4, GPT-3.5-turbo, and PALM2 in producing Verilog code for problems of different complexities?
% \item RQ6: At what levels of problem difficulty do the instructional LLMs excel, and where might they need enhancement to either meet or surpass the refined CodeGen-16B-FT model's performance?
% \item RQ7:  Are instructional LLMs like GPT4, GPT-3.5-turbo, and PALM2 sensitive to the problem description in generating Verilog code?
% \end{itemize}

We also conducted an evaluation of the fine-tuned models with the new \textit{large} LLMs such as OpenAI's state-of-the-art chatGPT (GPT-3.5-turbo)~\citep{ye2023comprehensive}, GPT4~\citep{openai_gpt-4_2023}, and Google's PALM2~\citep{chowdhery_palm_2022}. GPT4 posed a unique challenge due to its substantial costs and limited API access, which is closed-source and currently only accessible through a waitlist. To navigate around this constraint, we resorted to the web interface of GPT4. To facilitate this, we relied on a specific GitHub library\footnote{\url{https://github.com/acheong08/ChatGPT}}, which also allowed us to set the temperature at 0.2. This approach limited us to a single completion (n=1) per problem; thus, our evaluation process is framed accordingly. These constraints are represented visually in the subsequent~\autoref{fig:t-n-change}.

The evaluation considered the problem Set I described in~\autoref{tbl:problem_set-1} with varying complexity and prompt descriptions. We characterize the quality of compelted  Verilog by LLMs using Pass@(Scenario*$n$). That is, the number of solutions out of Scenariox$n$ on which the online system returns a \textit{Success}. Due to limited access to GPT4, we evaluate the LLMs using Pass@(Scenario*$n$) for $n=1$ completions per problem.

\subsubsection{Impact of variation in system prompt}\label{subsubsec:system-prompt}
Before we delve into the comparative results, let's first examine the significant impact of varying prompts on GPT-3.5-turbo. This model, unlike others, requires an additional input: the system prompt. This prompt essentially serves as an instruction set for the model and its influence on the resultant Verilog code was tested under two different scenarios.

In the first scenario, we used an \textit{unguided} system prompt that merely instructed the model to behave like a programming assistant, tasked with completing the code based on a user's description. The second scenario, however, involved a \textit{guided} system prompt. This prompt was far more specific, directing the model to function as a Verilog autocomplete engine, tasked with completing a partially written Verilog module, formatting the response appropriately, and ensuring the completion of the module at all times.

Following this adjustment, the GPT-3.5-turbo model was tested with both unguided (GPT-3.5-turbo-v0) and guided (GPT-3.5-turbo-v1) system prompts, as shown in \autoref{fig:t-n-change}.

In the Basic problem difficulty, the minimal prompt model surpasses the detailed prompt model by a staggering 53\%. This trend was consistent across problem difficulties from basic to advanced, and prompt descriptions from low (L) to high (H). The detailed prompt version consistently outperformed the minimal prompt version, demonstrating the influence of detailed instructions in the system prompt for generating high-quality Verilog code.

Looking at the prompt description level, the detailed prompt consistently outperformed the minimal prompt across all levels. Notably, at Low (L) description level, GPT-3.5-turbo-v1 performed approximately 41\% better than GPT-3.5-turbo-v0, and at High (H) description level, GPT-3.5-turbo-v1 again significantly outperformed GPT-3.5-turbo-v0, showing an improvement of approximately 34.3\%. These numbers clearly highlight that a richer, more detailed prompt greatly enhances the model's capability to generate high-quality Verilog code across various difficulties and description levels.

\subsubsection{Results of comparison with models under study}
From the LLMs evaluated in previous Section~\ref{sec:Results}, the fine-tuned CodeGen-16B-FT model, despite having only 16 billion parameters, solves up to 74\% of medium complexity problems when compared to the state-of-the-art model gpt-3.5-turbo and gpt4 which solves upto 75\% (approximate).

% In our evaluation of LLMs in Section~\ref{sec:Results}, the CodeGen-16B-FT model has displayed impressive performance. 
Despite its smaller size compared to state-of-the-art models like GPT4 and GPT-3.5-turbo, it showed proficiency across different problem description levels. The CodeGen-16B-FT model outperformed both GPT4 and GPT-3.5-turbo on tasks with medium-description (M) prompts, achieving a score of 0.436 compared to GPT4's 0.39 and GPT-3.5-turbo's 0.40. PALM2, on the other hand, shows a performance drops as the complexity increases, indicating potential areas for improvement.

 When compared to the large LLMs evaluated in ~\autoref{fig:t-n-change}, the fine-tuned CodeGen-16B-FT model exhibits commendable performance across varying problem complexities. In terms of the average problem-solving score, the \textit{large} LLMs such as GPT4, GPT-3.5-turbo, and PALM2 managed scores of approximately 0.53, 0.41, and 0.30 (approx.) respectively. In comparison, the CodeGen-16B-FT model demonstrated a competitive average score of 0.40 (approx.).[\textbf{Ans. RQ5}]\\

The performance metrics saw a shift with the introduction of larger LLMs like GPT4 and GPT-3.5-turbo, with parameter counts in the trillions. GPT4, in particular, showed proficiency in advanced problem-solving tasks, consistently scoring 0.6 across all levels of prompt detail, with a slight dip in intermediate-level problems, which is lower than the CodeGen-16B-FT's score of 0.54 (approximate). Despite this, an element of uncertainty remains due to their commercial/blackbox nature, which could impact their long-term availability.

While some \textit{large} LLMs, such as GPT4, excel in certain areas (like advanced problem-solving), they do not necessarily outperform in all contexts. Its performance is equivalent to that of CodeGen-16B-FT in intermediate-level problems, where it scored an average score of 0.75 (approximate).This indicates the need for more nuanced and detailed instructions to guide these models, especially for complex problems.[\textbf{Ans. RQ6}]\\

% While it did not surpass GPT4 on advanced problems, it demonstrated solid performance, particularly in basic problem-solving scenarios where it scored as high as 0.74.
Despite the entry of larger and advanced LLMs, the fine-tuned CodeGen-16B-FT model maintained its reliability across all problem difficulties. As we continue to evaluate and compare LLMs, it is evident that each has its strengths and areas for improvement. GPT4 excels in advanced problems, while a fine-tuned CodeGen-16B-FT showcases good performance across problem complexities; further fine-tuning with a diverse and well represented Verilog problems will improve its capability.

\subsection{Evaluation on expanded problem set}

\begin{figure}[H]
    \centering
    \includegraphics[width=0.78\linewidth]{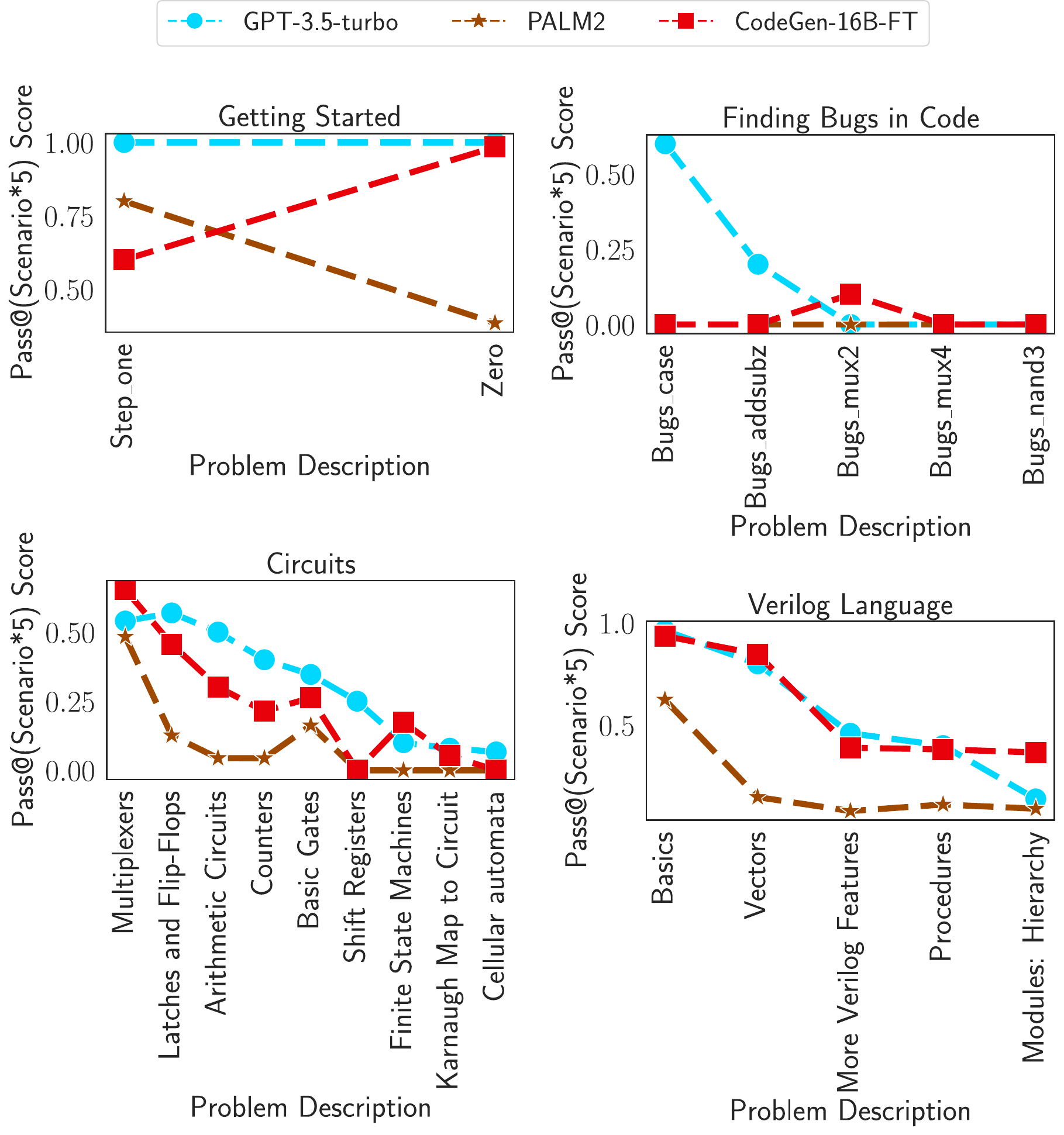}
    % \vspace{-1.5em}
    \caption{Pass@(scenario*n) at n=5 for scenarios passing test benches across categories on problem Set II (~\autoref{tbl:problems-set-2}). }
    \label{fig:hdlproblems-result}
\end{figure}

Using the extended problem set, we compare the quality of Verilog quality generated using three LLMs for Verilog: CodeGen-16B-FT, GPT-3.5-turbo, and PALM2. 
We use the Pass@($k$) metric to evaluate each model, where $k$ represents the total number of completions resulting from $\textit{Scenario} \times n$. Here, the term \textit{Scenario} refers to the combination of problems within a category, while $n$ is the number of completions produced per problem. Given the substantial costs associated with querying such a broad spectrum of problems across the LLM, such as GPT-3.5-turbo, we confined our inquiries to those with a temperature setting of 0.2. We generated five completions per problem across all categories. Each problem was scored on its ability to pass the extensive test benches provided by HDLBits, which include a rigorous suite of functional tests to ensure the correctness of the generated Verilog code.

\textit{Circuits}: The results from Figure~\ref{fig:hdlproblems-result} suggest varying degrees of proficiency across the models and problem categories. CodeGen-16B-FT outperforms the others in handling complex \textit{Circuits} problems, particularly those related to \textit{Multiplexers}, scoring a Pass@(Scenario*$n$) Score of 0.653. This is a significant improvement over GPT-3.5-turbo and PALM2, which scored 0.54 and 0.483, respectively, on similar problems.

\textit{Getting Started}: However, GPT-3.5-turbo demonstrates remarkable proficiency in \textit{Getting Started} problems, scoring a perfect 1.0 in the \textit{Step\_one} and \textit{Zero} categories, suggesting a strong grasp of basic Verilog concepts. In comparison, CodeGen-16B-FT is not far behind, while PALM2 lags noticeably behind in these foundational tasks.

\textit{Verification: Reading Simulations \& Finding Bugs in Code}: This category comprises bug fixing problems poses a considerable challenge to all three models. GPT-3.5-turbo achieves a top score of 0.6, but CodeGen-16B-FT had a noticeable edge over the others in tasks related to \textit{BugMux2}. PALM2, on the other hand, trails behind with a score of 0.42, indicating room for improvement in its ability to interpret and work with Verilog simulations.

\textit{Verilog Language}: In this category, CodeGen-16B-FT showed a strong understanding of \textit{Basics} and \textit{Vectors}, closely following GPT-3.5-turbo's scores. Interestingly, it also exceeded the performance of PALM2 in \textit{More Verilog Features}.

We were unable to include GPT-4 due to limited API access. Evaluating GPT-4 using a web interface for such a wide array of problems is challenging and future work will include GPT-4, once access limitations are eased.

\section{Impact of Training Data on Verilog Quality}

\begin{figure}[htp]
    \centering
    \begin{minipage}{0.39\textwidth}
        \centering

        \begin{forest}
        for tree={
            draw, % This line adds a border around each node
            align=center,
            l=2cm
        }
        [CodeGen-2B
            [CodeGen-2B-FT*, label=below:(a), edge label={node[midway,fill=white, font=\scriptsize]{Book Corpora}}]
            [CodeGen-2B-FT, edge label={node[midway,fill=white, font=\scriptsize]{Verilog </>}}
                [CodeGen-2B-FT++, label=below:(b), edge label={node[midway,fill=white, font=\scriptsize]{Book Corpus}}]
            ]
        ]
        \end{forest}

        \caption{Illustration of fine-tuning: a) CodeGen-2B-FT* is fine-tuned on Book Corpora, b) CodeGen-2B-FT fine-tuned on Verilog, followed by fine-tuned on Book Corpora yielding CodeGen-2B-FT++.}

    \end{minipage}
    \hspace*{-1in}
    \hfill
     \begin{minipage}{0.6\textwidth}
        \centering
        \includegraphics[width=0.85\linewidth]{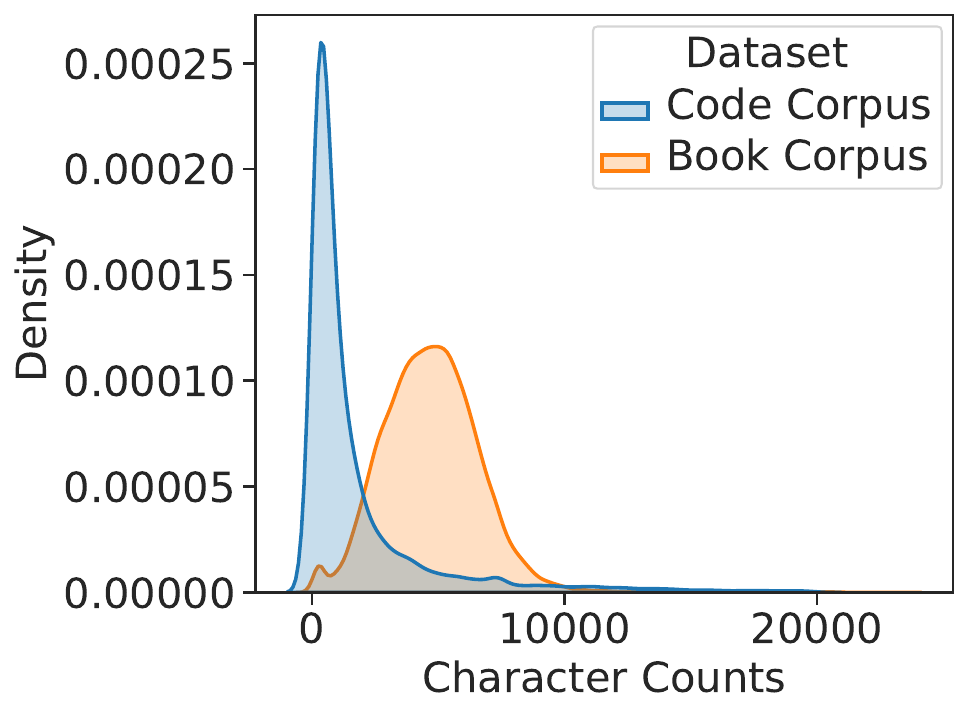} % second figure itself
        \caption{Distribution of character count for Verilog code  and book corpus}
        \label{fig:data-dist}
    \end{minipage}
\end{figure}

In this study, we examined the impact of training corpus content on Verilog code generation quality using the CodeGen-2B model. We compared two fine-tuning methodologies: (a) using only textbook Verilog content (CodeGen-2B-FT*), and (b) combining GitHub Verilog code and textbook content (CodeGen-2B-FT++). Our aim was to assess the effect of textbook-based Verilog content on LLM performance evaluated on problem Set I via the Pass@(scenario*10) metric with ten completions per problem.
%We demonstrate results from an experiment in 

Figure~\ref{fig:desc-diff-book} shows the Pass@(Scenario*$n$) score across  problem difficulties and descriptions. CodeGen-2B-FT++, which was fine-tuned using Verilog code and textbook content, showed significant improvements in Verilog code generation compared to CodeGen-2B-FT and CodeGen-2B-FT*.

Figure~\ref{fig:desc-diff-book} also shows that CodeGen-2B-FT++,  consistently outperformed the other LLMs across problem difficulty and description detail. In low (L) description problems, CodeGen-2B-FT++ showed a 10\% improvement over CodeGen-2B-FT. The improvement is more noticeable when compared to CodeGen-2B-FT*, which scored only 0.083 in the same category. This trend continues with medium (M) and high (H) description problems.

For basic problem-solving tasks, CodeGen-2B-FT++ achieved a Pass@(Scenario*10) score of 0.548, only slightly better than CodeGen-2B-FT, which scored 0.547. However, the performance leap is significant when compared to CodeGen-2B-FT*, which scored only 0.077 for basic tasks. This trend continues across intermediate and advanced problems and across low (L). CodeGen-2B-FT++ consistently outperforms the other variants of the model across all the levels of problem descriptions. 

This impressive performance of CodeGen-2B-FT++ can be attributed to the textbook content in the training data. Textbooks provide an array of examples, detailed explanations, and a broader context for Verilog use. This enriched training data allowed the LLM to form a more comprehensive understanding of Verilog, leading to the generation of not just correct but idiomatic and well-structured Verilog code.[\textbf{Ans. RQ7}]\\
An illustration of a problem to count from 1 to 12 is shown in Figure~\ref{fig:comp-book-counter} that points to the superiority of CodeGen-16B-FT++ over other variants. The task required a counter in Verilog that incrementally counts from 1 to 12 on the positive edge of the clock signal, resetting to 1 after reaching 12.

CodeGen-16B-FT*, trained solely on Verilog content from textbooks, failed to generate functionally correct code. It erroneously introduced an initial block, which is not meant for sequential logic, resulting in non-synthesizable code. Moreover, it ignored the requirement to reset the counter to 1 after reaching 12, thereby lacking proper functionality.
%On the other hand, 
CodeGen-16B-FT, trained only on Verilog code, produced code with the correct sequential logic structure, using an \texttt{always @(posedge clk)} block. However, it failed to meet the task's requirement, as it did not implement the wrap-around  from 12 back to 1. The code increments the \texttt{q} value without any check for resetting it back to 1 when it reaches 12. By contrast, CodeGen-16B-FT++, which had been fine-tuned using a mix of Verilog code and textbook content, successfully generated functionally correct and well-structured Verilog code. This model grasped the task requirements,  implementing the counter functionality with a wrap-around from 12 to 1. The resulting code uses an \texttt{always @(posedge clk)} block for sequential logic and includes conditions for resetting back to 1 on reaching 12.

The results of this study highlight the importance of diversifying the training corpus for large language models, with evidence showing marked improvement in model performance when a combination of practical code examples and educational resources is used.

\begin{figure}[th]
    \centering
    \includegraphics[width=0.92\linewidth]{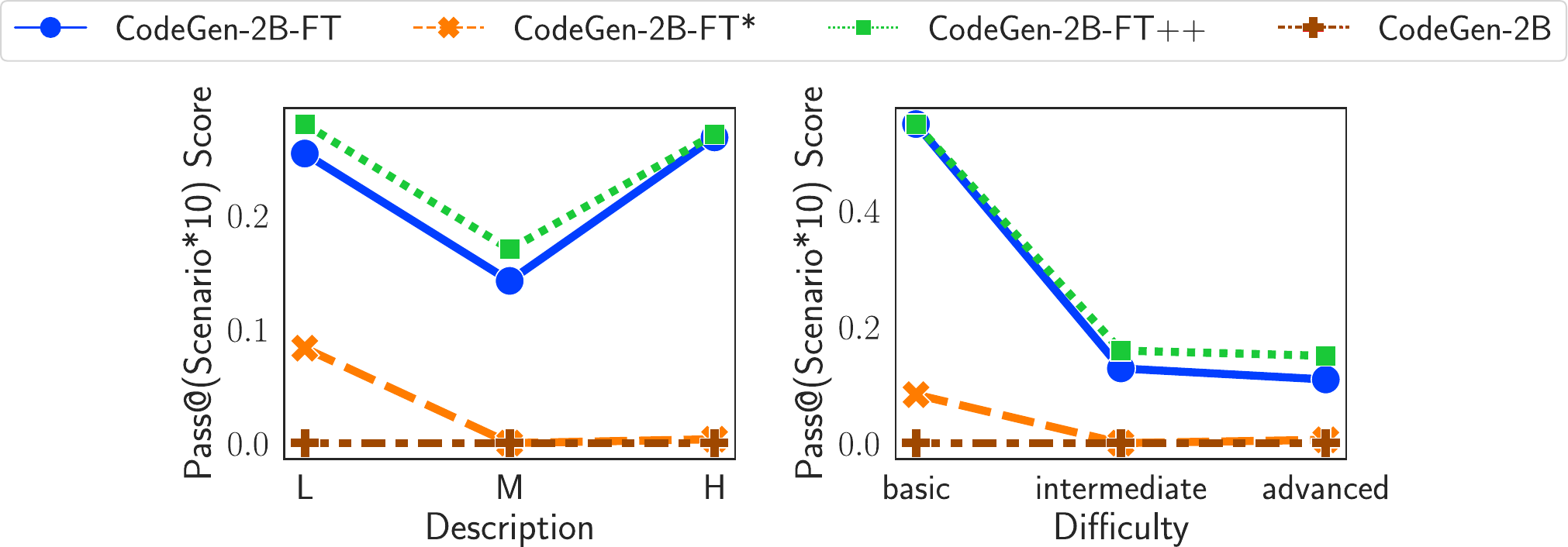}
    % \vspace{-1.5em}
    \caption{Pass@(scenario*10) at $n=10$ for scenarios passing test-benches on problem Set I across CodeGen-16B-FT* , CodeGen-16B-FT++, and CodeGen-16B-FT}
    \label{fig:desc-diff-book}
\end{figure}

\begin{figure}[h]
    \centering
    \includegraphics[width=0.99\linewidth]{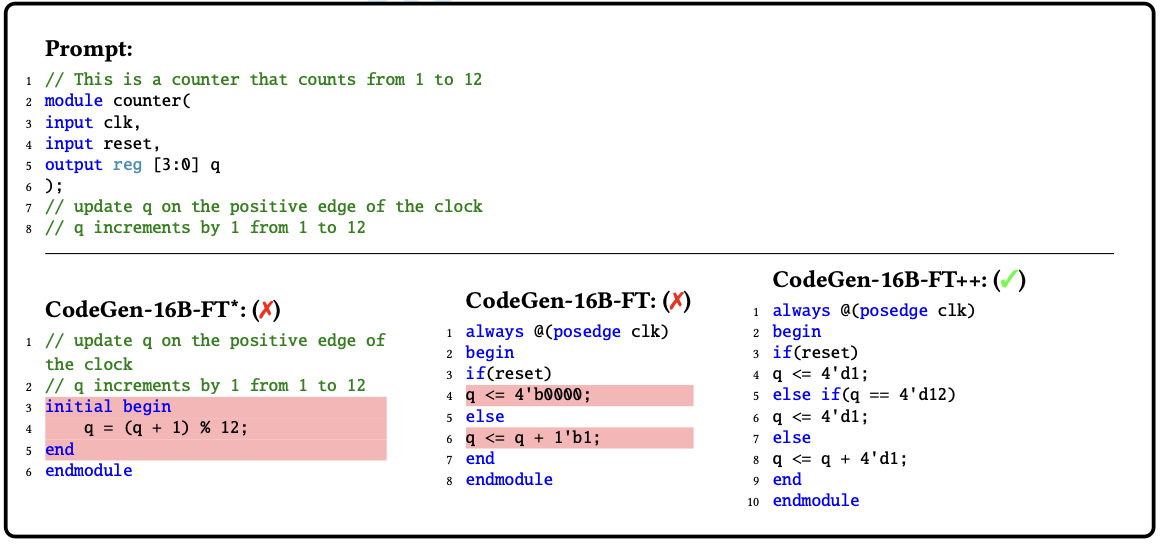}
    
\caption{CodeGen-16B-FT* vs CodeGen-16B-FT++ vs CodeGen-16B-FT at implementing Counter 1-12. We highlight the \fcolorbox{aired}{aired}{mistake}. }
\label{fig:comp-book-counter}
\end{figure}

\subsection{Inference Time Evaluation}

\begin{figure}[H]
    \centering
    \includegraphics[width=0.68\linewidth]{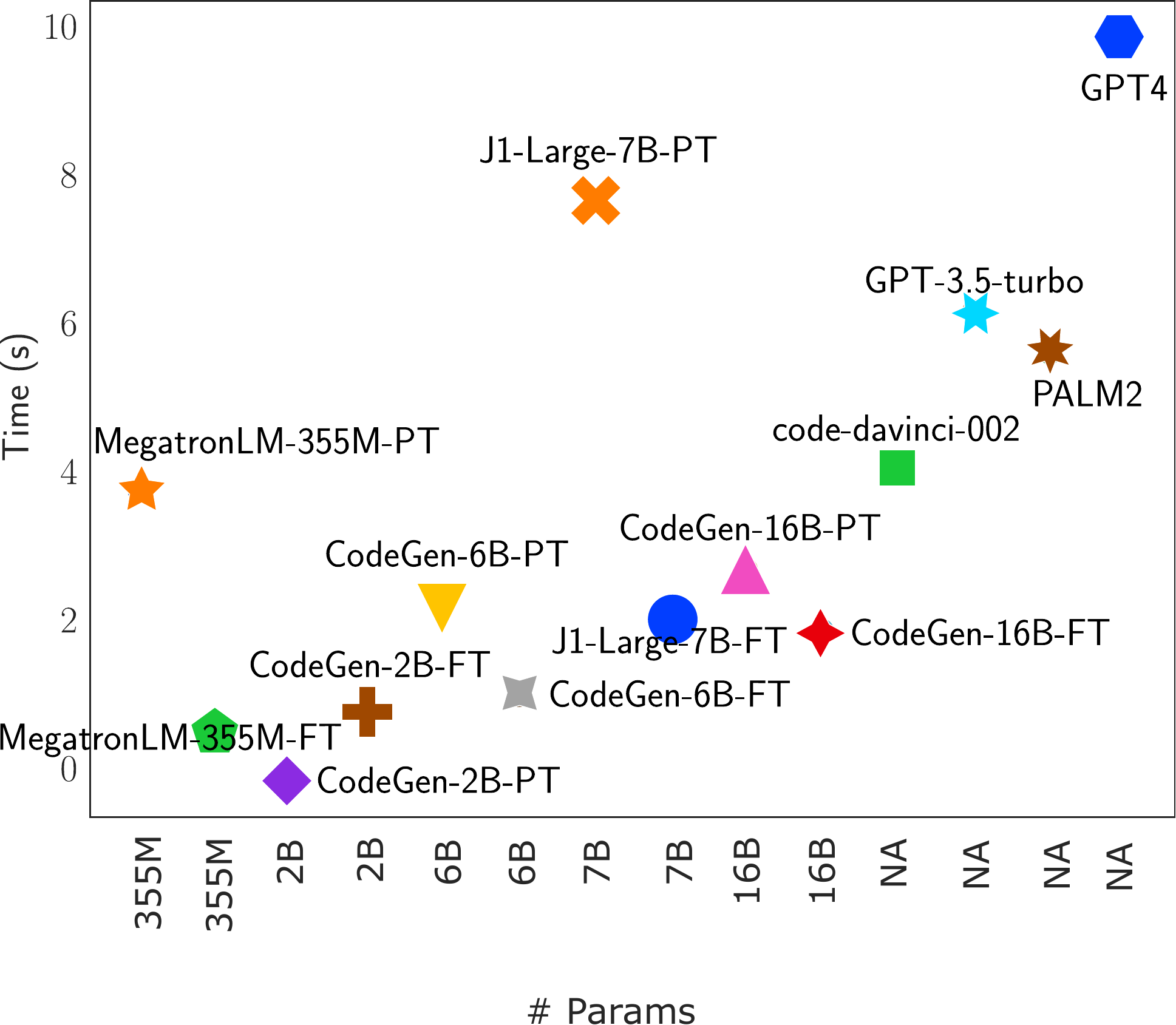}
    % \vspace{-1.5em}
    \caption{Inference time (s) for different LLMs. The \# of parameters  for PALM2, GPT-3.5-turbo, and GPT4 are not available (NA).}
    \label{fig:inference-results}
\end{figure}

Inference time plays a crucial role in the real-world application of language models, particularly when generating Verilog code for hardware design, where timeliness is often as important as the quality of the generated code. The following evaluation considers the response time in seconds of the models under consideration.

Figure~\ref{fig:inference-results} shows a scatter plot of the inference time in seconds (s) across the LLMs, where inference time is the total time taken from the time a prompt is fed into the model to the time the complete response is generated by the respective LLMs. Our locally fine-tuned model, CodeGen-16B-FT, demonstrates an efficient balance between response time and code quality. With an inference time of 2.02 seconds, it performs well against larger models like GPT-3.5-turbo, PALM2, and GPT4, which have inference times of 6.32, 5.76, and 10.000 seconds respectively. Despite these models showing competitive performance in generating high-quality Verilog code, their comparatively longer response times can pose a challenge for practical applications where rapid feedback is essential.

%However, it's worth noting that 
Smaller LLMs, CodeGen-2B-FT and CodeGen-6B-FT have inference times of 0.92 and 1.14 seconds, respectively. While their Verilog code generation capabilities may not match that of CodeGen-16B-FT, fast response times coupled with reasonable performance make them valuable assets in the initial stages of hardware design. A hardware designer may prefer these LLMs to generate a quick first-pass  code, providing a starting point that can then be refined according to  specific requirements.
The CodeGen-2B-PT model demonstrates the lowest inference time of all models, at 0.000 seconds. While this might imply an attractive level of efficiency, it is critical to consider this in conjunction with its problem-solving abilities, which are outperformed by our fine-tuned LLMs. Models such as J1-Large-7B-FT, despite a decent inference time of 2.12 seconds, do not quite match the Verilog code generation performance of CodeGen-16B-FT model, underscoring the necessity of balancing response times with  problem-solving capabilities.

%In conclusion, 
The fine-tuned LLMs, particularly CodeGen-16B-FT, exhibit a compelling combination of response time and code quality, making them attractive choices for real-world Verilog code generation tasks. However, smaller models like CodeGen-2B-FT and CodeGen-6B-FT, with their fast response times, can provide valuable first-pass code generation as part of an efficient hardware design workflow. These findings highlight the value of fine-tuning to  balance performance in  inference time and the quality of generated code.
\end{review}
\section{Discussion and Limitations}
\label{sec:discussion}

\subsection{Discussion of Example Scenarios}
\begin{review}

% significance of higher compilation rate: Enables designer to worry about syntax less.
Fine-tuned LLMs generate code that compiles better when compared to the pre-trained LLMs (\autoref{tbl:results}). Using the best Pass@(\textit{scenario}*$10$) values, only $11.9\%$ of the completions generated by pre-trained LLMs compiled vs. $64.6\%$ of those by fine-tuned LLMs. % compiled. 
Thus, a designer may use these \acp{LLM} with text/pseudo-code to generate a syntactically-correct design ``skeleton'', and tweak it to meet functional requirements.

% sensitivity in test benches can affect whether an answer passes or not. Use FSM clockedge example 
% We assess \acp{LLM}' code completion using the associated Verilog test-benches.
We used test benches to assess the generated Verilog. 
%  put in the box
These test-benches are comprehensive for the Basic problems, but as the problems become more complex, the test-benches cover only those behaviors fully specified in the problem comments.
%Given the simple nature of the Basic 
%We hand-crafted the test benches for each problem so they do not follow a standard format, given the diversity of the problems. This means that the `strictness' of the test bench can impact whether code completions pass. 
As LLMs tend to provide similar responses when several completions per prompt are requested, the exact test-bench implementation can have a large impact on how many cases pass. 
We observe this in the \acp{LLM}' responses to FSM problems 8, 15, and 17. As the problem comments do not specify whether the reset is synchronous/asynchronous, the LLMs are free to produce any variation.
For all problems, we verify whether an active-high reset results in the correct value at the output, but we do not test the asynchronous/synchronous  and other corner cases. 

\begin{figure}[h]
    \centering
    \includegraphics[width=0.99\linewidth]{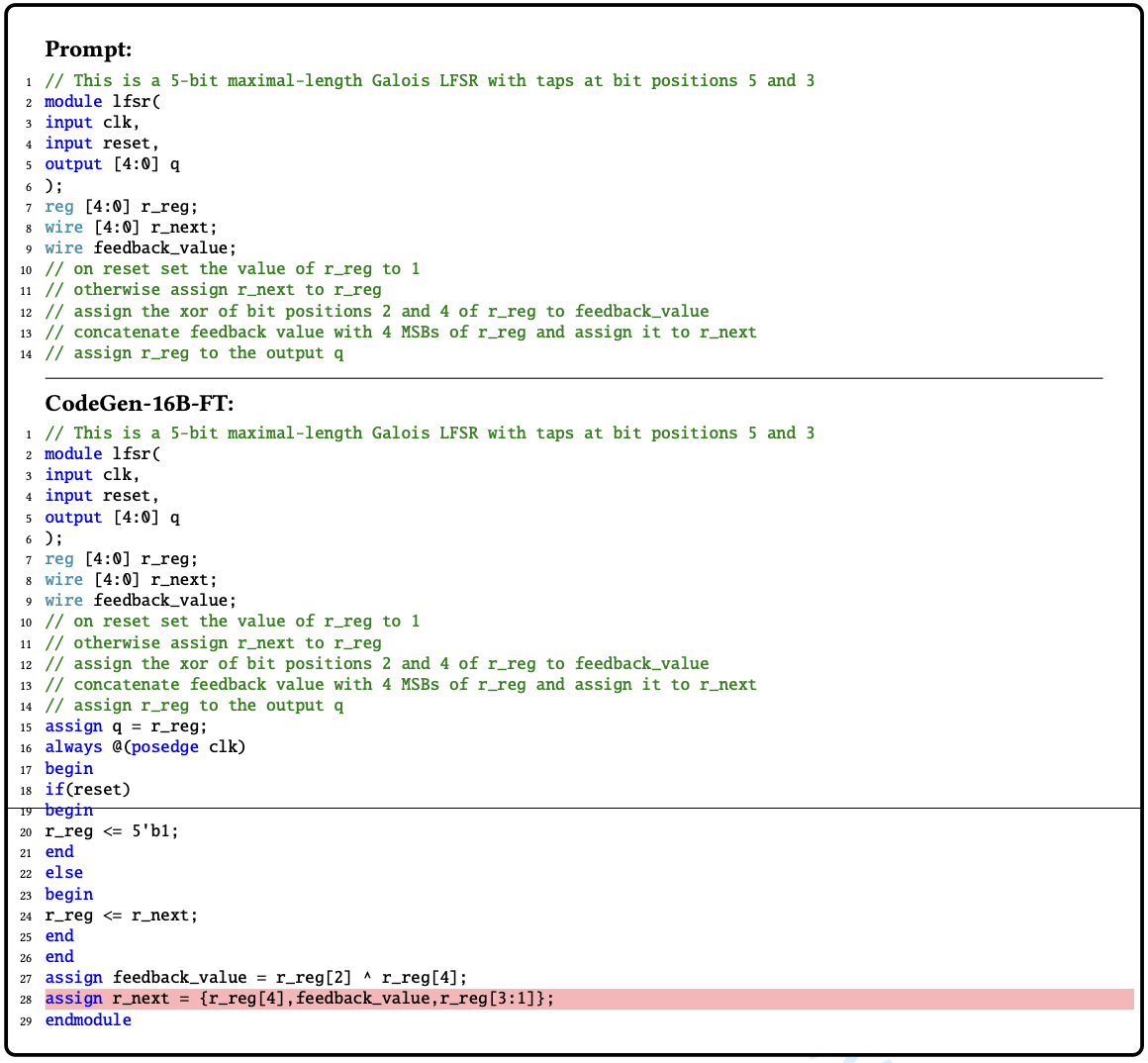}
    
\caption{Problem 7, Description: H, Difficulty Level: Intermediate~\autoref{tbl:problem_set-1}. We highlight the \fcolorbox{aired}{aired}{mistake}.}
\label{fig:comp-lfsr}
\end{figure}

\begin{figure}[h]
    \centering
    \includegraphics[width=0.99\linewidth]{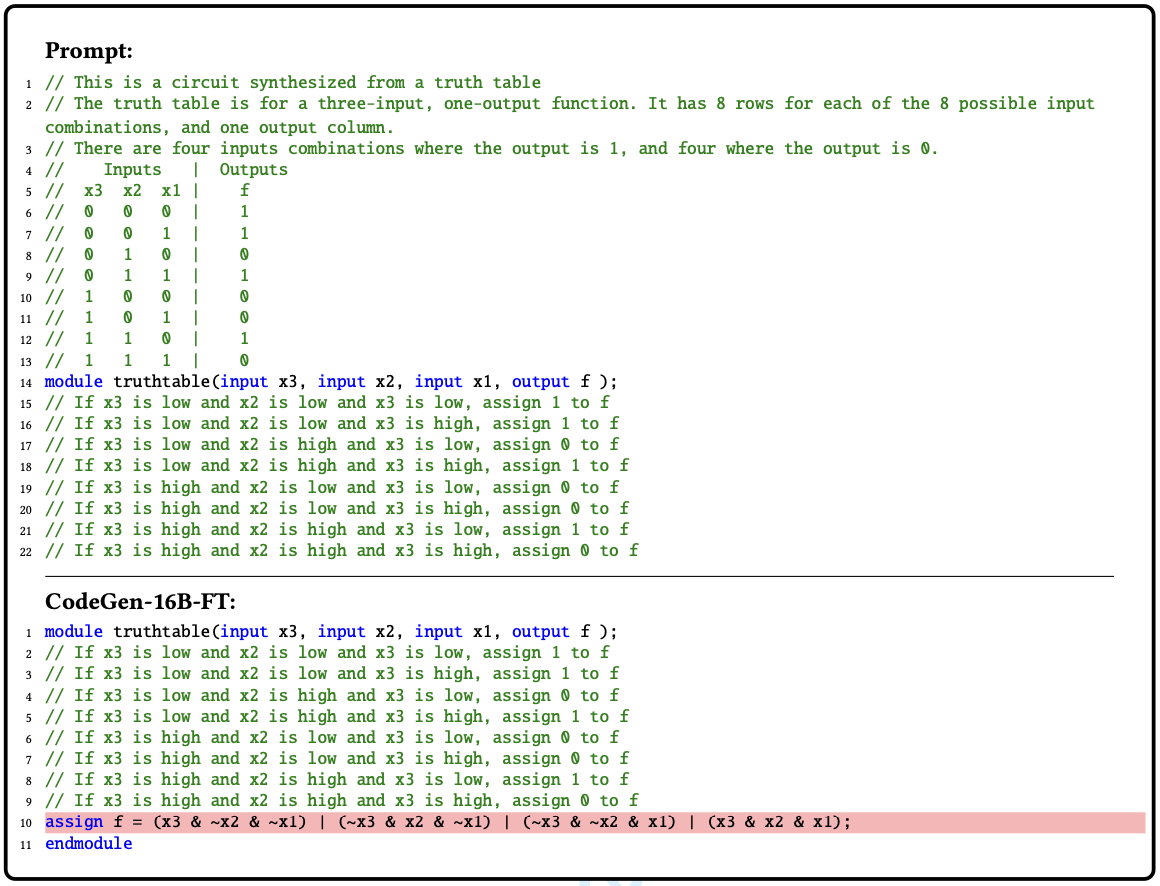}
    
\caption{Problem 12, Description: H, Difficulty Level: Intermediate~\autoref{tbl:problem_set-1}. We highlight the \fcolorbox{aired}{aired}{mistake}.}
\label{fig:comp-truthtable}
\end{figure}

% \begin{lrbox}{\mintedboxone}
% \begin{minipage}{\linewidth}
% {\bf CodeGen-16B-FT:}
% \begin{lstlisting}
% always @(posedge clk or posedge reset)
% begin
%         if(reset)
%         begin
%         out <= 0;
%         end
%         else
%         begin
%         if(load)
%         begin
%         out <= data;
%         end
%         else
%         begin
%         out <= {out[6:0],out[7]};
%         out[7:1] <= out[6:0];
%         end
%         end
% end
% endmodule
% \end{lstlisting}
% \end{minipage}
% \end{lrbox}

% \begin{figure}
% \begin{AIbox}
% \noindent\textbf{Prompt:}\begin{lstlisting}
% // This is a shift left and rotate operation
% module left_rotate(input clk,input reset,input [2:0] amount,input [7:0] data,input load,output reg [7:0] out);
% // when load is high, load data to out
% // when load is low, shift left and rotate the register out by amount bits
% \end{lstlisting}
% \par\vspace{0.1cm}\hrule\vspace{0.2cm}
% \usebox{\mintedboxone}
% \end{AIbox}
% \caption{Problem 9, Description: H, Difficulty Level: Intermediate,~\autoref{tbl:problem_set-1}. We highlight the \fcolorbox{aired}{aired}{mistake}.}
% \label{fig:comp-shiftrotate}
% \end{figure}

\begin{figure}[h]
    \centering
    \includegraphics[width=0.99\linewidth]{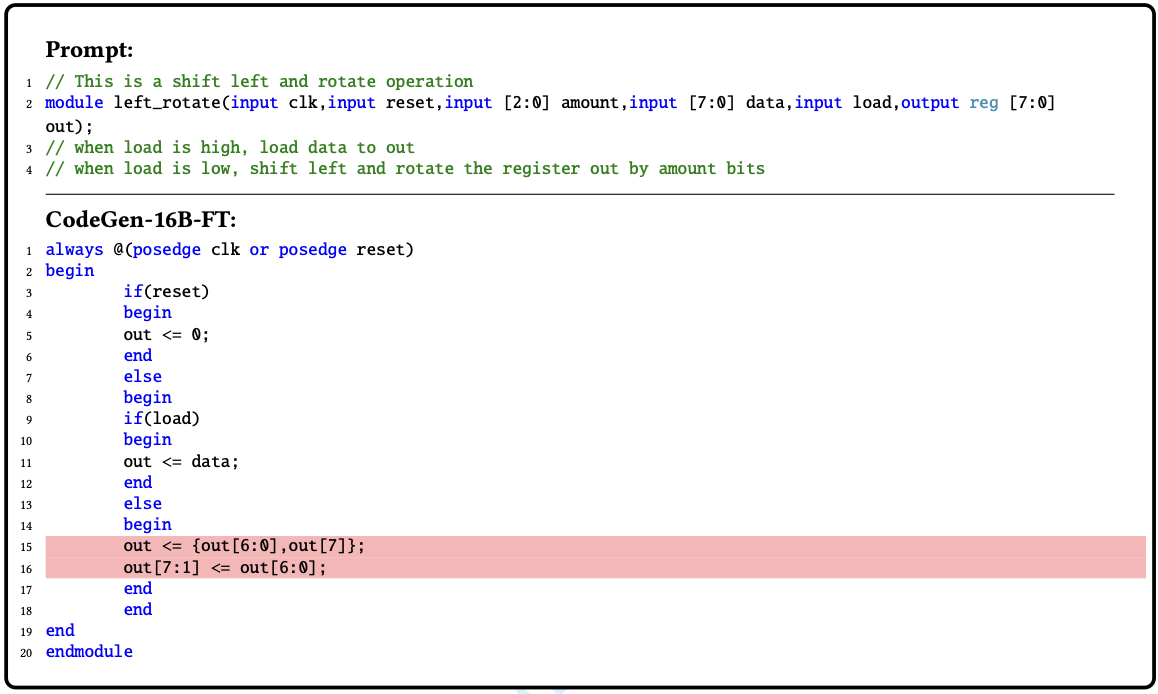}
    
\caption{Problem 9, Description: H, Difficulty Level: Intermediate,~\autoref{tbl:problem_set-1}. We highlight the \fcolorbox{aired}{aired}{mistake}.}
\label{fig:comp-shiftrotate}
\end{figure}

% Why did some problems perform so poorly?
The best-performing LLM (CodeGen-16B (FT)) performed poorly for some problem sets. %This trend was observed across the LLMs. 
For any given problem from problem Set I, CodeGen-16B (FT) produced 540 completions, but for Problems 7 (LFSR) in~\autoref{fig:comp-lfsr} and 12 (Truth table) in~\autoref{fig:comp-truthtable}, none passed, and for Problem 9 (Shift and Rotate) in~\autoref{fig:comp-shiftrotate}, only one passed. We inspected the completions and observed that for Prob. \# 7, the LLMs did not concatenate the most significant bits with the feedback value. %They overuse all bits instead of the 4 required bits. 
This was the problem in most cases, and a better prompt might yield a correct result, pointing to prompt engineering as future work.
For Prob. \#9, completions~\ref{fig:comp-shiftrotate}  either do not cover all values of the shift or assign incorrect bit positions. For Prob. \#12, completions~\ref{fig:comp-truthtable} are close to the actual solution by using all input values in \texttt{assign} statements but fail to form correct expressions between input bits. This suggests a lack of diversity in the training corpus.

We find that this trend extends to complex problems in problem Set II. The response to Conway's Game of Life problem, Category 14 from~\autoref{tbl:problems-set-2} shown in~\autoref{fig:comp-conway} illustrates this issue further. None of the LLMs -- GPT3.5-turbo, PALM2, or CodeGen-16B-FT - were able to generate code that achieved full functional correctness. They exhibit similar tendencies as observed in the FSM problems, often falling short in accurately handling edge cases and intricacies of the given problem.
The implementations produced for this problem by GPT-3.5-turbo and CodeGen-16B-FT shown in Figure~\ref{fig:comp-conway} had unique issues. 
In the first implementation, we encounter a  race condition as the model attempts to modify the \texttt{next\_state} and \texttt{current\_state} simultaneously at each clock edge, causing a conflict. The \texttt{next\_state} is declared as a wire, yet it is updated within an always block, which is not appropriate. In the second implementation, the game logic is flawed, as the ternary operation does not align with the Conway's Game of Life rules. Furthermore, the LLM fails to consider the toroidal nature of the game grid in the neighbor-counting function and mistakenly includes the cell itself as a neighbor.

This reinforces our observation that while LLMs have significantly improved in generating compilable "skeleton" code, achieving functionally correct code, especially in complex problems, continues to be a challenge. The generated code needs manual adjustments to handle edge cases and fully align with the problem's requirements. This makes the role of the designer crucial in refining the output of the LLMs to ensure functional correctness.

\begin{figure}[h]
    \centering
    \includegraphics[width=0.99\linewidth]{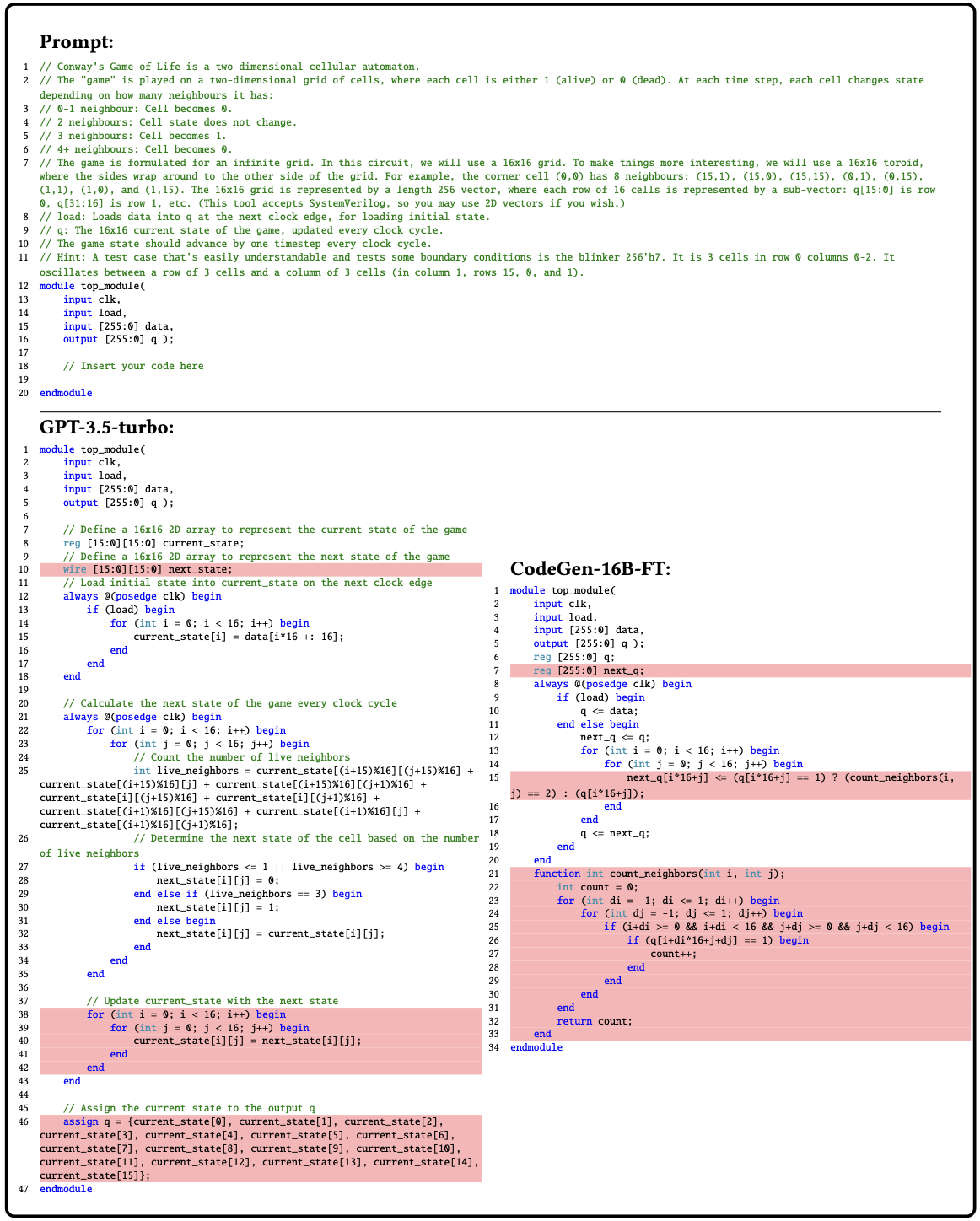}
    
\caption{GPT-3.5-turbo vs CodeGen-16B-FT at implementing Conway game of life problem, category: Cellular Automata, Table~\ref{tbl:problems-set-2}. We highlight the \fcolorbox{aired}{aired}{mistakes}. }
\label{fig:comp-conway}
\end{figure}

\begin{figure}[h]
    \centering
    \includegraphics[width=0.99\linewidth]{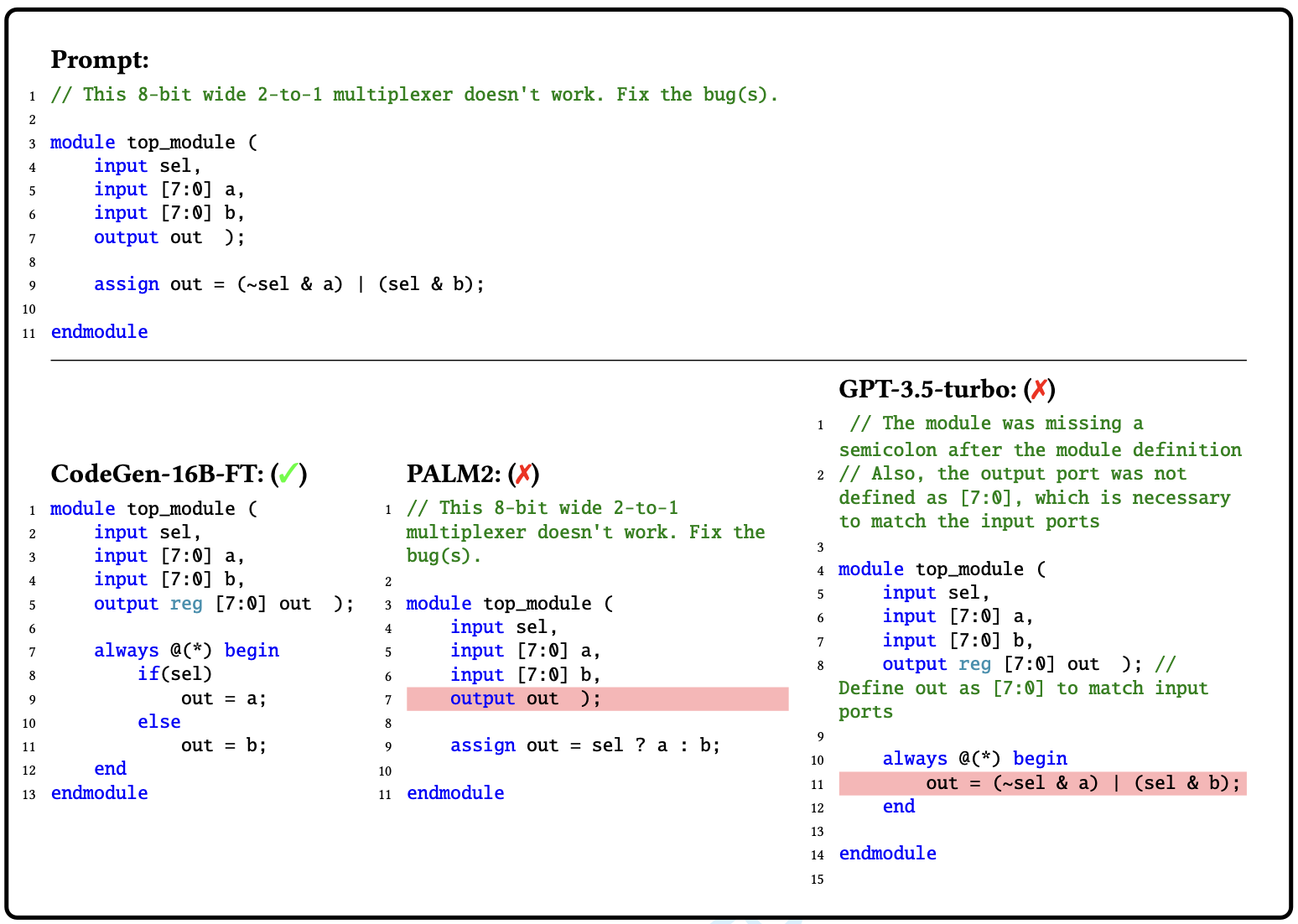}
    
\caption{CodeGen-16B-FT vs GPT-3.5-turbo vs PALM2 at fixing bug in 8-bit 2-to-1 MUX, category: Finding Bugs in Code,~\autoref{tbl:problems-set-2}. We highlight the \fcolorbox{aired}{aired}{mistake}. }
\label{fig:bug-mux2}
\end{figure}

While LLMs are trained on vast corpora of correct code, this doesn't inherently equip them with the ability to identify incorrect code and resolve the issues. Furthermore, the problem of error localization poses a unique challenge. For instance, we analyzed the bug-fixing capabilities of LLMs using a problem related to an 8-bit 2-to-1 multiplexer. As shown in the prompt description~\autoref{fig:bug-mux2}, the error was not localized and instead embedded subtly within the code's structure. This required the LLMs to have a deeper comprehension of the code beyond the local context, something which they currently struggle with. 

The solution generated by CodeGen-16B-FT is correct. It correctly identified and rectified the problem in the Verilog code. The second solution generated by PALM2 had a partial correction, but it still missed a crucial detail - the output port width. While the model correctly employed the ternary operator for the MUX operation, it failed to adjust the output port width to match the input, further reinforcing the issues seen in signal width handling. In the third solution, GPT-3.5-turbo correctly recognized the bug in the bitwise operations with different width signals. However, the solution did not fix the bug appropriately, implying a gap in the model's understanding of hardware operation nuances, particularly in interfacing signals of different bit widths.

\subsection{Conclusions}
This study introduces a novel framework for automatically creating and validating Verilog code using LLMs.
Drawing on the Pass@(scenario*$n$) values detailed in Tables~\ref{tbl:compiled}-\ref{tbl:results}, we found that pre-tuned LLMs only generated functionally accurate completions $1.09\%$ of the time. However, this figure jumps to $27.0\%$ post-tuning, demonstrating the significant advantages of fine-tuning LLMs to cater to a specific language. Refined CodeGen-16B LLM is the most proficient in generating functionally correct completions.
On the whole, it delivered functionally accurate code $41.9\%$ of the time, while the commercial top-tier (non-fine-tuned) code-davinci-002 LLM yielded functionally correct code $35.4\%$ of  time.

New entrant, GPT4 has shown outstanding performance across the problems, particularly it did well on advanced problem-solves. Smaller, fine-tuned LLMs like CodeGen-16B-FT are proficient and adaptable across a variety of problem complexities. Choosing an LLM is a balanced decision, considering use-case requirements, computational resources, problem complexity, cost, and reliability, particularly for commercial LLMs.

%As for CodeGen-16B-FT, 
Despite its impressive performance, there is room for improvement for CodeGen-16B-FT. Future work could refine its fine-tuning process, perhaps through targeted training data or additional iterations. Exploring hybrid approaches that combine the strengths of different LLMs is another potential avenue. There is the prospect of integrating domain-specific knowledge into the model to enhance its understanding of Verilog code generation. With these enhancements, CodeGen-16B-FT could compete with larger LLMs like GPT4 in terms of efficiency and output quality. %As the field continues to evolve, the potential for improvement and innovation remains vast.

\end{review}

\bibliographystyle{ACM-Reference-Format}
\bibliography{lit/benhamram,lit/new_citations}


\begin{thebibliography}{31}

%%% the Web address.

\ifx \showCODEN    \undefined \def \showCODEN     #1{\unskip}     \fi
\ifx \showDOI      \undefined \def \showDOI       #1{#1}\fi
\ifx \showISBNx    \undefined \def \showISBNx     #1{\unskip}     \fi
\ifx \showISBNxiii \undefined \def \showISBNxiii  #1{\unskip}     \fi
\ifx \showISSN     \undefined \def \showISSN      #1{\unskip}     \fi
\ifx \showLCCN     \undefined \def \showLCCN      #1{\unskip}     \fi
\ifx \shownote     \undefined \def \shownote      #1{#1}          \fi
\ifx \showarticletitle \undefined \def \showarticletitle #1{#1}   \fi
\ifx \showURL      \undefined \def \showURL       {\relax}        \fi
% invisible to TeX
\providecommand\bibfield[2]{#2}
\providecommand\bibinfo[2]{#2}
\providecommand\natexlab[1]{#1}
\providecommand\showeprint[2][]{arXiv:#2}

\bibitem[\protect\citeauthoryear{Ahmad, Waseem, Liang, Fehmideh, Aktar, and
  Mikkonen}{Ahmad et~al\mbox{.}}{2023}]%
        {ahmad_towards_2023}
\bibfield{author}{\bibinfo{person}{Aakash Ahmad}, \bibinfo{person}{Muhammad
  Waseem}, \bibinfo{person}{Peng Liang}, \bibinfo{person}{Mahdi Fehmideh},
  \bibinfo{person}{Mst~Shamima Aktar}, {and} \bibinfo{person}{Tommi Mikkonen}.}
  \bibinfo{year}{2023}\natexlab{}.
\newblock \bibinfo{title}{Towards {Human}-{Bot} {Collaborative} {Software}
  {Architecting} with {ChatGPT}}.
\newblock
\newblock
\urldef\tempurl%
\url{https://doi.org/10.48550/arXiv.2302.14600}
\showDOI{\tempurl}
\newblock
\shownote{arXiv:2302.14600 [cs].}


\bibitem[\protect\citeauthoryear{AI21}{AI21}{2021}]%
        {ai21_jurassic-1_2021}
\bibfield{author}{\bibinfo{person}{AI21}.} \bibinfo{year}{2021}\natexlab{}.
\newblock \bibinfo{title}{Jurassic-1 {Language} {Models} - {AI21} {Studio}
  {Docs}}.
\newblock
\newblock
\urldef\tempurl%
\url{https://studio.ai21.com/docs/jurassic1-language-models/#general-purpose-models}
\showURL{%
\tempurl}


\bibitem[\protect\citeauthoryear{Bachrach, Vo, Richards, Lee, Waterman, Avi{\v
  z}ienis, Wawrzynek, and Asanovi{\'c}}{Bachrach et~al\mbox{.}}{2012}]%
        {bachrach_chisel_2012}
\bibfield{author}{\bibinfo{person}{Jonathan Bachrach}, \bibinfo{person}{Huy
  Vo}, \bibinfo{person}{Brian Richards}, \bibinfo{person}{Yunsup Lee},
  \bibinfo{person}{Andrew Waterman}, \bibinfo{person}{Rimas Avi{\v z}ienis},
  \bibinfo{person}{John Wawrzynek}, {and} \bibinfo{person}{Krste
  Asanovi{\'c}}.} \bibinfo{year}{2012}\natexlab{}.
\newblock \showarticletitle{Chisel: constructing hardware in a {Scala} embedded
  language}. In \bibinfo{booktitle}{\emph{Proceedings of the 49th {Annual}
  {Design} {Automation} {Conference}}} \emph{(\bibinfo{series}{{DAC} '12})}.
  \bibinfo{publisher}{Association for Computing Machinery},
  \bibinfo{address}{New York, NY, USA}, \bibinfo{pages}{1216--1225}.
\newblock
\showISBNx{978-1-4503-1199-1}
\urldef\tempurl%
\url{https://doi.org/10.1145/2228360.2228584}
\showDOI{\tempurl}


\bibitem[\protect\citeauthoryear{Brown, Mann, Ryder, Subbiah, Kaplan, Dhariwal,
  Neelakantan, Shyam, Sastry, Askell, Agarwal, Herbert-Voss, Krueger, Henighan,
  Child, Ramesh, Ziegler, Wu, Winter, Hesse, Chen, Sigler, Litwin, Gray, Chess,
  Clark, Berner, McCandlish, Radford, Sutskever, and Amodei}{Brown
  et~al\mbox{.}}{2020}]%
        {brown_language_2020}
\bibfield{author}{\bibinfo{person}{Tom Brown}, \bibinfo{person}{Benjamin Mann},
  \bibinfo{person}{Nick Ryder}, \bibinfo{person}{Melanie Subbiah},
  \bibinfo{person}{Jared~D Kaplan}, \bibinfo{person}{Prafulla Dhariwal},
  \bibinfo{person}{Arvind Neelakantan}, \bibinfo{person}{Pranav Shyam},
  \bibinfo{person}{Girish Sastry}, \bibinfo{person}{Amanda Askell},
  \bibinfo{person}{Sandhini Agarwal}, \bibinfo{person}{Ariel Herbert-Voss},
  \bibinfo{person}{Gretchen Krueger}, \bibinfo{person}{Tom Henighan},
  \bibinfo{person}{Rewon Child}, \bibinfo{person}{Aditya Ramesh},
  \bibinfo{person}{Daniel Ziegler}, \bibinfo{person}{Jeffrey Wu},
  \bibinfo{person}{Clemens Winter}, \bibinfo{person}{Chris Hesse},
  \bibinfo{person}{Mark Chen}, \bibinfo{person}{Eric Sigler},
  \bibinfo{person}{Mateusz Litwin}, \bibinfo{person}{Scott Gray},
  \bibinfo{person}{Benjamin Chess}, \bibinfo{person}{Jack Clark},
  \bibinfo{person}{Christopher Berner}, \bibinfo{person}{Sam McCandlish},
  \bibinfo{person}{Alec Radford}, \bibinfo{person}{Ilya Sutskever}, {and}
  \bibinfo{person}{Dario Amodei}.} \bibinfo{year}{2020}\natexlab{}.
\newblock \showarticletitle{Language {Models} are {Few}-{Shot} {Learners}}. In
  \bibinfo{booktitle}{\emph{Advances in {Neural} {Information} {Processing}
  {Systems}}}, \bibfield{editor}{\bibinfo{person}{H.~Larochelle},
  \bibinfo{person}{M.~Ranzato}, \bibinfo{person}{R.~Hadsell},
  \bibinfo{person}{M.~F. Balcan}, {and} \bibinfo{person}{H.~Lin}} (Eds.),
  Vol.~\bibinfo{volume}{33}. \bibinfo{publisher}{Curran Associates, Inc.},
  \bibinfo{pages}{1877--1901}.
\newblock
\urldef\tempurl%
\url{https://proceedings.neurips.cc/paper/2020/file/1457c0d6bfcb4967418bfb8ac142f64a-Paper.pdf}
\showURL{%
\tempurl}


\bibitem[\protect\citeauthoryear{Chen, Tworek, Jun, Yuan, Pinto, Kaplan,
  Edwards, Burda, Joseph, Brockman, Ray, Puri, Krueger, Petrov, Khlaaf, Sastry,
  Mishkin, Chan, Gray, Ryder, Pavlov, Power, Kaiser, Bavarian, Winter, Tillet,
  Such, Cummings, Plappert, Chantzis, Barnes, Herbert-Voss, Guss, Nichol,
  Paino, Tezak, Tang, Babuschkin, Balaji, Jain, Saunders, Hesse, Carr, Leike,
  Achiam, Misra, Morikawa, Radford, Knight, Brundage, Murati, Mayer, Welinder,
  McGrew, Amodei, McCandlish, Sutskever, and Zaremba}{Chen
  et~al\mbox{.}}{2021}]%
        {chen_evaluating_2021}
\bibfield{author}{\bibinfo{person}{Mark Chen}, \bibinfo{person}{Jerry Tworek},
  \bibinfo{person}{Heewoo Jun}, \bibinfo{person}{Qiming Yuan},
  \bibinfo{person}{Henrique Ponde de~Oliveira Pinto}, \bibinfo{person}{Jared
  Kaplan}, \bibinfo{person}{Harri Edwards}, \bibinfo{person}{Yuri Burda},
  \bibinfo{person}{Nicholas Joseph}, \bibinfo{person}{Greg Brockman},
  \bibinfo{person}{Alex Ray}, \bibinfo{person}{Raul Puri},
  \bibinfo{person}{Gretchen Krueger}, \bibinfo{person}{Michael Petrov},
  \bibinfo{person}{Heidy Khlaaf}, \bibinfo{person}{Girish Sastry},
  \bibinfo{person}{Pamela Mishkin}, \bibinfo{person}{Brooke Chan},
  \bibinfo{person}{Scott Gray}, \bibinfo{person}{Nick Ryder},
  \bibinfo{person}{Mikhail Pavlov}, \bibinfo{person}{Alethea Power},
  \bibinfo{person}{Lukasz Kaiser}, \bibinfo{person}{Mohammad Bavarian},
  \bibinfo{person}{Clemens Winter}, \bibinfo{person}{Philippe Tillet},
  \bibinfo{person}{Felipe~Petroski Such}, \bibinfo{person}{Dave Cummings},
  \bibinfo{person}{Matthias Plappert}, \bibinfo{person}{Fotios Chantzis},
  \bibinfo{person}{Elizabeth Barnes}, \bibinfo{person}{Ariel Herbert-Voss},
  \bibinfo{person}{William~Hebgen Guss}, \bibinfo{person}{Alex Nichol},
  \bibinfo{person}{Alex Paino}, \bibinfo{person}{Nikolas Tezak},
  \bibinfo{person}{Jie Tang}, \bibinfo{person}{Igor Babuschkin},
  \bibinfo{person}{Suchir Balaji}, \bibinfo{person}{Shantanu Jain},
  \bibinfo{person}{William Saunders}, \bibinfo{person}{Christopher Hesse},
  \bibinfo{person}{Andrew~N. Carr}, \bibinfo{person}{Jan Leike},
  \bibinfo{person}{Josh Achiam}, \bibinfo{person}{Vedant Misra},
  \bibinfo{person}{Evan Morikawa}, \bibinfo{person}{Alec Radford},
  \bibinfo{person}{Matthew Knight}, \bibinfo{person}{Miles Brundage},
  \bibinfo{person}{Mira Murati}, \bibinfo{person}{Katie Mayer},
  \bibinfo{person}{Peter Welinder}, \bibinfo{person}{Bob McGrew},
  \bibinfo{person}{Dario Amodei}, \bibinfo{person}{Sam McCandlish},
  \bibinfo{person}{Ilya Sutskever}, {and} \bibinfo{person}{Wojciech Zaremba}.}
  \bibinfo{year}{2021}\natexlab{}.
\newblock \bibinfo{title}{Evaluating {Large} {Language} {Models} {Trained} on
  {Code}}.
\newblock
\newblock
\urldef\tempurl%
\url{https://doi.org/10.48550/arXiv.2107.03374}
\showDOI{\tempurl}
\newblock
\shownote{arXiv:2107.03374 [cs].}


\bibitem[\protect\citeauthoryear{Chowdhery, Narang, Devlin, Bosma, Mishra,
  Roberts, Barham, Chung, Sutton, Gehrmann, Schuh, Shi, Tsvyashchenko, Maynez,
  Rao, Barnes, Tay, Shazeer, Prabhakaran, Reif, Du, Hutchinson, Pope, Bradbury,
  Austin, Isard, Gur-Ari, Yin, Duke, Levskaya, Ghemawat, Dev, Michalewski,
  Garcia, Misra, Robinson, Fedus, Zhou, Ippolito, Luan, Lim, Zoph, Spiridonov,
  Sepassi, Dohan, Agrawal, Omernick, Dai, Pillai, Pellat, Lewkowycz, Moreira,
  Child, Polozov, Lee, Zhou, Wang, Saeta, Diaz, Firat, Catasta, Wei,
  Meier-Hellstern, Eck, Dean, Petrov, and Fiedel}{Chowdhery
  et~al\mbox{.}}{2022}]%
        {chowdhery_palm_2022}
\bibfield{author}{\bibinfo{person}{Aakanksha Chowdhery},
  \bibinfo{person}{Sharan Narang}, \bibinfo{person}{Jacob Devlin},
  \bibinfo{person}{Maarten Bosma}, \bibinfo{person}{Gaurav Mishra},
  \bibinfo{person}{Adam Roberts}, \bibinfo{person}{Paul Barham},
  \bibinfo{person}{Hyung~Won Chung}, \bibinfo{person}{Charles Sutton},
  \bibinfo{person}{Sebastian Gehrmann}, \bibinfo{person}{Parker Schuh},
  \bibinfo{person}{Kensen Shi}, \bibinfo{person}{Sasha Tsvyashchenko},
  \bibinfo{person}{Joshua Maynez}, \bibinfo{person}{Abhishek Rao},
  \bibinfo{person}{Parker Barnes}, \bibinfo{person}{Yi Tay},
  \bibinfo{person}{Noam Shazeer}, \bibinfo{person}{Vinodkumar Prabhakaran},
  \bibinfo{person}{Emily Reif}, \bibinfo{person}{Nan Du}, \bibinfo{person}{Ben
  Hutchinson}, \bibinfo{person}{Reiner Pope}, \bibinfo{person}{James Bradbury},
  \bibinfo{person}{Jacob Austin}, \bibinfo{person}{Michael Isard},
  \bibinfo{person}{Guy Gur-Ari}, \bibinfo{person}{Pengcheng Yin},
  \bibinfo{person}{Toju Duke}, \bibinfo{person}{Anselm Levskaya},
  \bibinfo{person}{Sanjay Ghemawat}, \bibinfo{person}{Sunipa Dev},
  \bibinfo{person}{Henryk Michalewski}, \bibinfo{person}{Xavier Garcia},
  \bibinfo{person}{Vedant Misra}, \bibinfo{person}{Kevin Robinson},
  \bibinfo{person}{Liam Fedus}, \bibinfo{person}{Denny Zhou},
  \bibinfo{person}{Daphne Ippolito}, \bibinfo{person}{David Luan},
  \bibinfo{person}{Hyeontaek Lim}, \bibinfo{person}{Barret Zoph},
  \bibinfo{person}{Alexander Spiridonov}, \bibinfo{person}{Ryan Sepassi},
  \bibinfo{person}{David Dohan}, \bibinfo{person}{Shivani Agrawal},
  \bibinfo{person}{Mark Omernick}, \bibinfo{person}{Andrew~M. Dai},
  \bibinfo{person}{Thanumalayan~Sankaranarayana Pillai}, \bibinfo{person}{Marie
  Pellat}, \bibinfo{person}{Aitor Lewkowycz}, \bibinfo{person}{Erica Moreira},
  \bibinfo{person}{Rewon Child}, \bibinfo{person}{Oleksandr Polozov},
  \bibinfo{person}{Katherine Lee}, \bibinfo{person}{Zongwei Zhou},
  \bibinfo{person}{Xuezhi Wang}, \bibinfo{person}{Brennan Saeta},
  \bibinfo{person}{Mark Diaz}, \bibinfo{person}{Orhan Firat},
  \bibinfo{person}{Michele Catasta}, \bibinfo{person}{Jason Wei},
  \bibinfo{person}{Kathy Meier-Hellstern}, \bibinfo{person}{Douglas Eck},
  \bibinfo{person}{Jeff Dean}, \bibinfo{person}{Slav Petrov}, {and}
  \bibinfo{person}{Noah Fiedel}.} \bibinfo{year}{2022}\natexlab{}.
\newblock \bibinfo{title}{{PaLM}: {Scaling} {Language} {Modeling} with
  {Pathways}}.
\newblock
\newblock
\urldef\tempurl%
\url{https://doi.org/10.48550/arXiv.2204.02311}
\showDOI{\tempurl}
\newblock
\shownote{arXiv:2204.02311 [cs].}


\bibitem[\protect\citeauthoryear{Dessouky, Gens, Haney, Persyn, Kanuparthi,
  Khattri, Fung, Sadeghi, and Rajendran}{Dessouky et~al\mbox{.}}{2019}]%
        {dessouky_hardfails_2019}
\bibfield{author}{\bibinfo{person}{Ghada Dessouky}, \bibinfo{person}{David
  Gens}, \bibinfo{person}{Patrick Haney}, \bibinfo{person}{Garrett Persyn},
  \bibinfo{person}{Arun Kanuparthi}, \bibinfo{person}{Hareesh Khattri},
  \bibinfo{person}{Jason~M. Fung}, \bibinfo{person}{Ahmad-Reza Sadeghi}, {and}
  \bibinfo{person}{Jeyavijayan Rajendran}.} \bibinfo{year}{2019}\natexlab{}.
\newblock \showarticletitle{{HardFails}: {Insights} into
  {Software}-{Exploitable} {Hardware} {Bugs}}. \bibinfo{pages}{213--230}.
\newblock
\showISBNx{978-1-939133-06-9}
\urldef\tempurl%
\url{https://www.usenix.org/conference/usenixsecurity19/presentation/dessouky}
\showURL{%
\tempurl}


\bibitem[\protect\citeauthoryear{Devlin, Chang, Lee, and Toutanova}{Devlin
  et~al\mbox{.}}{2019}]%
        {devlin_bert_2019}
\bibfield{author}{\bibinfo{person}{Jacob Devlin}, \bibinfo{person}{Ming-Wei
  Chang}, \bibinfo{person}{Kenton Lee}, {and} \bibinfo{person}{Kristina
  Toutanova}.} \bibinfo{year}{2019}\natexlab{}.
\newblock \showarticletitle{{BERT}: {Pre}-training of {Deep} {Bidirectional}
  {Transformers} for {Language} {Understanding}}. In
  \bibinfo{booktitle}{\emph{Proceedings of the 2019 {Conference} of the {North}
  {American} {Chapter} of the {Association} for {Computational} {Linguistics}:
  {Human} {Language} {Technologies}, {Volume} 1 ({Long} and {Short}
  {Papers})}}. \bibinfo{publisher}{Association for Computational Linguistics},
  \bibinfo{address}{Minneapolis, Minnesota}, \bibinfo{pages}{4171--4186}.
\newblock
\urldef\tempurl%
\url{https://doi.org/10.18653/v1/N19-1423}
\showDOI{\tempurl}


\bibitem[\protect\citeauthoryear{Drechsler, Harris, and Wille}{Drechsler
  et~al\mbox{.}}{2012}]%
        {drechsler_generating_2012}
\bibfield{author}{\bibinfo{person}{Rolf Drechsler}, \bibinfo{person}{Ian~G.
  Harris}, {and} \bibinfo{person}{Robert Wille}.}
  \bibinfo{year}{2012}\natexlab{}.
\newblock \showarticletitle{Generating formal system models from natural
  language descriptions}. In \bibinfo{booktitle}{\emph{{IEEE} {Int}. {High}
  {Level} {Design} {Validation} and {Test} {Workshop} ({HLDVT})}}.
  \bibinfo{pages}{164--165}.
\newblock


\bibitem[\protect\citeauthoryear{Gage}{Gage}{1994}]%
        {gage_new_1994}
\bibfield{author}{\bibinfo{person}{Philip Gage}.}
  \bibinfo{year}{1994}\natexlab{}.
\newblock \showarticletitle{A {New} {Algorithm} for {Data} {Compression}}.
\newblock \bibinfo{journal}{\emph{C Users Journal}} \bibinfo{volume}{12},
  \bibinfo{number}{2} (\bibinfo{date}{Feb.} \bibinfo{year}{1994}),
  \bibinfo{pages}{23--38}.
\newblock
\showISSN{0898-9788}


\bibitem[\protect\citeauthoryear{Gao, Biderman, Black, Golding, Hoppe, Foster,
  Phang, He, Thite, Nabeshima, Presser, and Leahy}{Gao et~al\mbox{.}}{2020}]%
        {gao_pile_2020}
\bibfield{author}{\bibinfo{person}{Leo Gao}, \bibinfo{person}{Stella Biderman},
  \bibinfo{person}{Sid Black}, \bibinfo{person}{Laurence Golding},
  \bibinfo{person}{Travis Hoppe}, \bibinfo{person}{Charles Foster},
  \bibinfo{person}{Jason Phang}, \bibinfo{person}{Horace He},
  \bibinfo{person}{Anish Thite}, \bibinfo{person}{Noa Nabeshima},
  \bibinfo{person}{Shawn Presser}, {and} \bibinfo{person}{Connor Leahy}.}
  \bibinfo{year}{2020}\natexlab{}.
\newblock \bibinfo{title}{The {Pile}: {An} {800GB} {Dataset} of {Diverse}
  {Text} for {Language} {Modeling}}.
\newblock
\newblock
\urldef\tempurl%
\url{https://doi.org/10.48550/arXiv.2101.00027}
\showDOI{\tempurl}
\newblock
\shownote{Number: arXiv:2101.00027 arXiv:2101.00027 [cs].}


\bibitem[\protect\citeauthoryear{Harris and Harris}{Harris and Harris}{2016}]%
        {harris_glast_2016}
\bibfield{author}{\bibinfo{person}{Christopher~B. Harris} {and}
  \bibinfo{person}{Ian~G. Harris}.} \bibinfo{year}{2016}\natexlab{}.
\newblock \showarticletitle{{GLAsT}: {Learning} formal grammars to translate
  natural language specifications into hardware assertions}. In
  \bibinfo{booktitle}{\emph{Design, {Automation} {Test} in {Europe} {Conf}.
  {Exhibition} ({DATE})}}. \bibinfo{pages}{966--971}.
\newblock


\bibitem[\protect\citeauthoryear{Husain, Wu, Gazit, Allamanis, and
  Brockschmidt}{Husain et~al\mbox{.}}{2020}]%
        {codesearchnet}
\bibfield{author}{\bibinfo{person}{Hamel Husain}, \bibinfo{person}{Ho-Hsiang
  Wu}, \bibinfo{person}{Tiferet Gazit}, \bibinfo{person}{Miltiadis Allamanis},
  {and} \bibinfo{person}{Marc Brockschmidt}.} \bibinfo{year}{2020}\natexlab{}.
\newblock \bibinfo{title}{CodeSearchNet Challenge: Evaluating the State of
  Semantic Code Search}.
\newblock
\newblock
\showeprint[arxiv]{1909.09436}~[cs.LG]


\bibitem[\protect\citeauthoryear{Li, Allal, Zi, Muennighoff, Kocetkov, Mou,
  Marone, Akiki, Li, Chim, Liu, Zheltonozhskii, Zhuo, Wang, Dehaene, Davaadorj,
  Lamy-Poirier, Monteiro, Shliazhko, Gontier, Meade, Zebaze, Yee, Umapathi,
  Zhu, Lipkin, Oblokulov, Wang, Murthy, Stillerman, Patel, Abulkhanov, Zocca,
  Dey, Zhang, Fahmy, Bhattacharyya, Yu, Singh, Luccioni, Villegas, Kunakov,
  Zhdanov, Romero, Lee, Timor, Ding, Schlesinger, Schoelkopf, Ebert, Dao,
  Mishra, Gu, Robinson, Anderson, Dolan-Gavitt, Contractor, Reddy, Fried,
  Bahdanau, Jernite, Ferrandis, Hughes, Wolf, Guha, von Werra, and de~Vries}{Li
  et~al\mbox{.}}{2023}]%
        {li_starcoder_2023}
\bibfield{author}{\bibinfo{person}{Raymond Li}, \bibinfo{person}{Loubna~Ben
  Allal}, \bibinfo{person}{Yangtian Zi}, \bibinfo{person}{Niklas Muennighoff},
  \bibinfo{person}{Denis Kocetkov}, \bibinfo{person}{Chenghao Mou},
  \bibinfo{person}{Marc Marone}, \bibinfo{person}{Christopher Akiki},
  \bibinfo{person}{Jia Li}, \bibinfo{person}{Jenny Chim}, \bibinfo{person}{Qian
  Liu}, \bibinfo{person}{Evgenii Zheltonozhskii}, \bibinfo{person}{Terry~Yue
  Zhuo}, \bibinfo{person}{Thomas Wang}, \bibinfo{person}{Olivier Dehaene},
  \bibinfo{person}{Mishig Davaadorj}, \bibinfo{person}{Joel Lamy-Poirier},
  \bibinfo{person}{Jo{\~a}o Monteiro}, \bibinfo{person}{Oleh Shliazhko},
  \bibinfo{person}{Nicolas Gontier}, \bibinfo{person}{Nicholas Meade},
  \bibinfo{person}{Armel Zebaze}, \bibinfo{person}{Ming-Ho Yee},
  \bibinfo{person}{Logesh~Kumar Umapathi}, \bibinfo{person}{Jian Zhu},
  \bibinfo{person}{Benjamin Lipkin}, \bibinfo{person}{Muhtasham Oblokulov},
  \bibinfo{person}{Zhiruo Wang}, \bibinfo{person}{Rudra Murthy},
  \bibinfo{person}{Jason Stillerman}, \bibinfo{person}{Siva~Sankalp Patel},
  \bibinfo{person}{Dmitry Abulkhanov}, \bibinfo{person}{Marco Zocca},
  \bibinfo{person}{Manan Dey}, \bibinfo{person}{Zhihan Zhang},
  \bibinfo{person}{Nour Fahmy}, \bibinfo{person}{Urvashi Bhattacharyya},
  \bibinfo{person}{Wenhao Yu}, \bibinfo{person}{Swayam Singh},
  \bibinfo{person}{Sasha Luccioni}, \bibinfo{person}{Paulo Villegas},
  \bibinfo{person}{Maxim Kunakov}, \bibinfo{person}{Fedor Zhdanov},
  \bibinfo{person}{Manuel Romero}, \bibinfo{person}{Tony Lee},
  \bibinfo{person}{Nadav Timor}, \bibinfo{person}{Jennifer Ding},
  \bibinfo{person}{Claire Schlesinger}, \bibinfo{person}{Hailey Schoelkopf},
  \bibinfo{person}{Jan Ebert}, \bibinfo{person}{Tri Dao},
  \bibinfo{person}{Mayank Mishra}, \bibinfo{person}{Alex Gu},
  \bibinfo{person}{Jennifer Robinson}, \bibinfo{person}{Carolyn~Jane Anderson},
  \bibinfo{person}{Brendan Dolan-Gavitt}, \bibinfo{person}{Danish Contractor},
  \bibinfo{person}{Siva Reddy}, \bibinfo{person}{Daniel Fried},
  \bibinfo{person}{Dzmitry Bahdanau}, \bibinfo{person}{Yacine Jernite},
  \bibinfo{person}{Carlos~Mu{\~n}oz Ferrandis}, \bibinfo{person}{Sean Hughes},
  \bibinfo{person}{Thomas Wolf}, \bibinfo{person}{Arjun Guha},
  \bibinfo{person}{Leandro von Werra}, {and} \bibinfo{person}{Harm de Vries}.}
  \bibinfo{year}{2023}\natexlab{}.
\newblock \bibinfo{title}{{StarCoder}: may the source be with you!}
\newblock
\newblock
\urldef\tempurl%
\url{https://doi.org/10.48550/arXiv.2305.06161}
\showDOI{\tempurl}
\newblock
\shownote{arXiv:2305.06161 [cs].}


\bibitem[\protect\citeauthoryear{Mihalcea, Liu, and Lieberman}{Mihalcea
  et~al\mbox{.}}{2006}]%
        {mihalcea_nlp_2006}
\bibfield{author}{\bibinfo{person}{Rada Mihalcea}, \bibinfo{person}{Hugo Liu},
  {and} \bibinfo{person}{Henry Lieberman}.} \bibinfo{year}{2006}\natexlab{}.
\newblock \showarticletitle{{NLP} ({Natural} {Language} {Processing}) for {NLP}
  ({Natural} {Language} {Programming})}. In
  \bibinfo{booktitle}{\emph{Computational {Linguistics} and {Intelligent}
  {Text} {Processing}}}, \bibfield{editor}{\bibinfo{person}{Alexander Gelbukh}}
  (Ed.). \bibinfo{publisher}{Springer Berlin Heidelberg},
  \bibinfo{pages}{319--330}.
\newblock
\showISBNx{978-3-540-32206-1}


\bibitem[\protect\citeauthoryear{Nijkamp, Pang, Hayashi, Tu, Wang, Zhou,
  Savarese, and Xiong}{Nijkamp et~al\mbox{.}}{2022}]%
        {nijkamp_conversational_2022}
\bibfield{author}{\bibinfo{person}{Erik Nijkamp}, \bibinfo{person}{Bo Pang},
  \bibinfo{person}{Hiroaki Hayashi}, \bibinfo{person}{Lifu Tu},
  \bibinfo{person}{Huan Wang}, \bibinfo{person}{Yingbo Zhou},
  \bibinfo{person}{Silvio Savarese}, {and} \bibinfo{person}{Caiming Xiong}.}
  \bibinfo{year}{2022}\natexlab{}.
\newblock \bibinfo{title}{A {Conversational} {Paradigm} for {Program}
  {Synthesis}}.
\newblock
\newblock
\urldef\tempurl%
\url{https://doi.org/10.48550/arXiv.2203.13474}
\showDOI{\tempurl}
\newblock
\shownote{arXiv:2203.13474 [cs].}


\bibitem[\protect\citeauthoryear{OpenAI}{OpenAI}{2023}]%
        {openai_gpt-4_2023}
\bibfield{author}{\bibinfo{person}{OpenAI}.} \bibinfo{year}{2023}\natexlab{}.
\newblock \bibinfo{title}{{GPT}-4}.
\newblock
\newblock
\urldef\tempurl%
\url{https://openai.com/research/gpt-4}
\showURL{%
\tempurl}


\bibitem[\protect\citeauthoryear{Pearce, Ahmad, Tan, Dolan-Gavitt, and
  Karri}{Pearce et~al\mbox{.}}{2022}]%
        {pearce_asleep_2022}
\bibfield{author}{\bibinfo{person}{Hammond Pearce}, \bibinfo{person}{Baleegh
  Ahmad}, \bibinfo{person}{Benjamin Tan}, \bibinfo{person}{Brendan
  Dolan-Gavitt}, {and} \bibinfo{person}{Ramesh Karri}.}
  \bibinfo{year}{2022}\natexlab{}.
\newblock \showarticletitle{Asleep at the {Keyboard}? {Assessing} the
  {Security} of {GitHub} {Copilot}{\textquoteright}s {Code} {Contributions}}.
  In \bibinfo{booktitle}{\emph{2022 {IEEE} {Symposium} on {Security} and
  {Privacy} ({SP})}}. \bibinfo{pages}{754--768}.
\newblock
\urldef\tempurl%
\url{https://doi.org/10.1109/SP46214.2022.9833571}
\showDOI{\tempurl}
\newblock
\shownote{ISSN: 2375-1207.}


\bibitem[\protect\citeauthoryear{Pearce, Tan, and Karri}{Pearce
  et~al\mbox{.}}{2020}]%
        {pearce_dave_2020}
\bibfield{author}{\bibinfo{person}{Hammond Pearce}, \bibinfo{person}{Benjamin
  Tan}, {and} \bibinfo{person}{Ramesh Karri}.} \bibinfo{year}{2020}\natexlab{}.
\newblock \showarticletitle{{DAVE}: {Deriving} {Automatically} {Verilog} from
  {English}}. In \bibinfo{booktitle}{\emph{Proceedings of the 2020 {ACM}/{IEEE}
  {Workshop} on {Machine} {Learning} for {CAD}}}. \bibinfo{publisher}{ACM},
  \bibinfo{address}{Virtual Event Iceland}, \bibinfo{pages}{27--32}.
\newblock
\showISBNx{978-1-4503-7519-1}
\urldef\tempurl%
\url{https://doi.org/10.1145/3380446.3430634}
\showDOI{\tempurl}


\bibitem[\protect\citeauthoryear{Radford, Wu, Child, Luan, Amodei, and
  Sutskever}{Radford et~al\mbox{.}}{2019}]%
        {radford_language_2019}
\bibfield{author}{\bibinfo{person}{Alec Radford}, \bibinfo{person}{Jeffrey Wu},
  \bibinfo{person}{Rewon Child}, \bibinfo{person}{David Luan},
  \bibinfo{person}{Dario Amodei}, {and} \bibinfo{person}{Ilya Sutskever}.}
  \bibinfo{year}{2019}\natexlab{}.
\newblock \showarticletitle{Language {Models} are {Unsupervised} {Multitask}
  {Learners}}.
\newblock  (\bibinfo{year}{2019}), \bibinfo{pages}{24}.
\newblock
\urldef\tempurl%
\url{https://cdn.openai.com/better-language-models/language_models_are_unsupervised_multitask_learners.pdf}
\showURL{%
\tempurl}


\bibitem[\protect\citeauthoryear{Rajbhandari, Ruwase, Rasley, Smith, and
  He}{Rajbhandari et~al\mbox{.}}{2021}]%
        {rajbhandari_zero-infinity_2021}
\bibfield{author}{\bibinfo{person}{Samyam Rajbhandari},
  \bibinfo{person}{Olatunji Ruwase}, \bibinfo{person}{Jeff Rasley},
  \bibinfo{person}{Shaden Smith}, {and} \bibinfo{person}{Yuxiong He}.}
  \bibinfo{year}{2021}\natexlab{}.
\newblock \bibinfo{title}{{ZeRO}-{Infinity}: {Breaking} the {GPU} {Memory}
  {Wall} for {Extreme} {Scale} {Deep} {Learning}}.
\newblock
\newblock
\urldef\tempurl%
\url{https://doi.org/10.48550/arXiv.2104.07857}
\showDOI{\tempurl}
\newblock
\shownote{Number: arXiv:2104.07857 arXiv:2104.07857 [cs].}


\bibitem[\protect\citeauthoryear{Rasley, Rajbhandari, Ruwase, and He}{Rasley
  et~al\mbox{.}}{2020}]%
        {deepspeed}
\bibfield{author}{\bibinfo{person}{Jeff Rasley}, \bibinfo{person}{Samyam
  Rajbhandari}, \bibinfo{person}{Olatunji Ruwase}, {and}
  \bibinfo{person}{Yuxiong He}.} \bibinfo{year}{2020}\natexlab{}.
\newblock \showarticletitle{DeepSpeed: System Optimizations Enable Training
  Deep Learning Models with Over 100 Billion Parameters}. In
  \bibinfo{booktitle}{\emph{Proceedings of the 26th ACM SIGKDD International
  Conference on Knowledge Discovery \& Data Mining}} (Virtual Event, CA, USA)
  \emph{(\bibinfo{series}{KDD '20})}. \bibinfo{publisher}{Association for
  Computing Machinery}, \bibinfo{address}{New York, NY, USA},
  \bibinfo{pages}{3505–3506}.
\newblock
\showISBNx{9781450379984}
\urldef\tempurl%
\url{https://doi.org/10.1145/3394486.3406703}
\showDOI{\tempurl}


\bibitem[\protect\citeauthoryear{Ren, Rajbhandari, Aminabadi, Ruwase, Yang,
  Zhang, Li, and He}{Ren et~al\mbox{.}}{2021}]%
        {ren_zero-offload_2021}
\bibfield{author}{\bibinfo{person}{Jie Ren}, \bibinfo{person}{Samyam
  Rajbhandari}, \bibinfo{person}{Reza~Yazdani Aminabadi},
  \bibinfo{person}{Olatunji Ruwase}, \bibinfo{person}{Shuangyan Yang},
  \bibinfo{person}{Minjia Zhang}, \bibinfo{person}{Dong Li}, {and}
  \bibinfo{person}{Yuxiong He}.} \bibinfo{year}{2021}\natexlab{}.
\newblock \bibinfo{title}{{ZeRO}-{Offload}: {Democratizing} {Billion}-{Scale}
  {Model} {Training}}.
\newblock
\newblock
\urldef\tempurl%
\url{https://doi.org/10.48550/arXiv.2101.06840}
\showDOI{\tempurl}
\newblock
\shownote{Number: arXiv:2101.06840 arXiv:2101.06840 [cs].}


\bibitem[\protect\citeauthoryear{Sandoval, Pearce, Nys, Karri, Garg, and
  Dolan-Gavitt}{Sandoval et~al\mbox{.}}{2023}]%
        {sandoval_lost_2023}
\bibfield{author}{\bibinfo{person}{Gustavo Sandoval}, \bibinfo{person}{Hammond
  Pearce}, \bibinfo{person}{Teo Nys}, \bibinfo{person}{Ramesh Karri},
  \bibinfo{person}{Siddharth Garg}, {and} \bibinfo{person}{Brendan
  Dolan-Gavitt}.} \bibinfo{year}{2023}\natexlab{}.
\newblock \showarticletitle{Lost at {C}: {A} {User} {Study} on the {Security}
  {Implications} of {Large} {Language} {Model} {Code} {Assistants}}.
\newblock \bibinfo{journal}{\emph{USENIX Security Symposium}}
  (\bibinfo{year}{2023}).
\newblock


\bibitem[\protect\citeauthoryear{Shoeybi, Patwary, Puri, LeGresley, Casper, and
  Catanzaro}{Shoeybi et~al\mbox{.}}{2020}]%
        {shoeybi_megatron-lm_2020}
\bibfield{author}{\bibinfo{person}{Mohammad Shoeybi}, \bibinfo{person}{Mostofa
  Patwary}, \bibinfo{person}{Raul Puri}, \bibinfo{person}{Patrick LeGresley},
  \bibinfo{person}{Jared Casper}, {and} \bibinfo{person}{Bryan Catanzaro}.}
  \bibinfo{year}{2020}\natexlab{}.
\newblock \bibinfo{title}{Megatron-{LM}: {Training} {Multi}-{Billion}
  {Parameter} {Language} {Models} {Using} {Model} {Parallelism}}.
\newblock
\newblock
\urldef\tempurl%
\url{http://arxiv.org/abs/1909.08053}
\showURL{%
\tempurl}
\newblock
\shownote{arXiv:1909.08053 [cs].}


\bibitem[\protect\citeauthoryear{Thakur, Ahmad, Fan, Pearce, Tan, Karri,
  Dolan-Gavitt, and Garg}{Thakur et~al\mbox{.}}{2023}]%
        {thakur_benchmarking_2023}
\bibfield{author}{\bibinfo{person}{Shailja Thakur}, \bibinfo{person}{Baleegh
  Ahmad}, \bibinfo{person}{Zhenxing Fan}, \bibinfo{person}{Hammond Pearce},
  \bibinfo{person}{Benjamin Tan}, \bibinfo{person}{Ramesh Karri},
  \bibinfo{person}{Brendan Dolan-Gavitt}, {and} \bibinfo{person}{Siddharth
  Garg}.} \bibinfo{year}{2023}\natexlab{}.
\newblock \showarticletitle{Benchmarking {Large} {Language} {Models} for
  {Automated} {Verilog} {RTL} {Code} {Generation}}. In
  \bibinfo{booktitle}{\emph{2023 {Design}, {Automation} \& {Test} in {Europe}
  {Conference} \& {Exhibition} ({DATE})}}. \bibinfo{pages}{1--6}.
\newblock
\urldef\tempurl%
\url{https://doi.org/10.23919/DATE56975.2023.10137086}
\showDOI{\tempurl}
\newblock
\shownote{ISSN: 1558-1101.}


\bibitem[\protect\citeauthoryear{Vaswani, Shazeer, Parmar, Uszkoreit, Jones,
  Gomez, Kaiser, and Polosukhin}{Vaswani et~al\mbox{.}}{2017}]%
        {vaswani_attention_2017}
\bibfield{author}{\bibinfo{person}{Ashish Vaswani}, \bibinfo{person}{Noam
  Shazeer}, \bibinfo{person}{Niki Parmar}, \bibinfo{person}{Jakob Uszkoreit},
  \bibinfo{person}{Llion Jones}, \bibinfo{person}{Aidan~N Gomez},
  \bibinfo{person}{{\L }ukasz Kaiser}, {and} \bibinfo{person}{Illia
  Polosukhin}.} \bibinfo{year}{2017}\natexlab{}.
\newblock \showarticletitle{Attention is {All} you {Need}}. In
  \bibinfo{booktitle}{\emph{Advances in {Neural} {Information} {Processing}
  {Systems}}}, Vol.~\bibinfo{volume}{30}. \bibinfo{publisher}{Curran
  Associates, Inc.}
\newblock
\urldef\tempurl%
\url{https://proceedings.neurips.cc/paper/2017/hash/3f5ee243547dee91fbd053c1c4a845aa-Abstract.html}
\showURL{%
\tempurl}


\bibitem[\protect\citeauthoryear{Williams}{Williams}{2023}]%
        {williams_icarus_2023}
\bibfield{author}{\bibinfo{person}{Stephen Williams}.}
  \bibinfo{year}{2023}\natexlab{}.
\newblock \bibinfo{title}{The {ICARUS} {Verilog} {Compilation} {System}}.
\newblock
\newblock
\urldef\tempurl%
\url{https://github.com/steveicarus/iverilog}
\showURL{%
\tempurl}
\newblock
\shownote{original-date: 2008-05-12T16:57:52Z.}


\bibitem[\protect\citeauthoryear{Wong}{Wong}{2019}]%
        {wong_projectabout_2019}
\bibfield{author}{\bibinfo{person}{Henry Wong}.}
  \bibinfo{year}{2019}\natexlab{}.
\newblock \bibinfo{title}{Project:{About} - {HDLBits}}.
\newblock
\newblock
\urldef\tempurl%
\url{https://hdlbits.01xz.net/wiki/Project:About}
\showURL{%
\tempurl}


\bibitem[\protect\citeauthoryear{Yan, Liu, Li, Han, and Qiu}{Yan
  et~al\mbox{.}}{2017}]%
        {yan_privmin_2017}
\bibfield{author}{\bibinfo{person}{Ziqi Yan}, \bibinfo{person}{Jiqiang Liu},
  \bibinfo{person}{Gang Li}, \bibinfo{person}{Zhen Han}, {and}
  \bibinfo{person}{Shuo Qiu}.} \bibinfo{year}{2017}\natexlab{}.
\newblock \bibinfo{title}{{PrivMin}: {Differentially} {Private} {MinHash} for
  {Jaccard} {Similarity} {Computation}}.
\newblock
\newblock
\urldef\tempurl%
\url{https://doi.org/10.48550/arXiv.1705.07258}
\showDOI{\tempurl}
\newblock
\shownote{Number: arXiv:1705.07258 arXiv:1705.07258 [cs].}


\bibitem[\protect\citeauthoryear{Ye, Chen, Xu, Zu, Shao, Liu, Cui, Zhou, Gong,
  Shen, Zhou, Chen, Gui, Zhang, and Huang}{Ye et~al\mbox{.}}{2023}]%
        {ye2023comprehensive}
\bibfield{author}{\bibinfo{person}{Junjie Ye}, \bibinfo{person}{Xuanting Chen},
  \bibinfo{person}{Nuo Xu}, \bibinfo{person}{Can Zu}, \bibinfo{person}{Zekai
  Shao}, \bibinfo{person}{Shichun Liu}, \bibinfo{person}{Yuhan Cui},
  \bibinfo{person}{Zeyang Zhou}, \bibinfo{person}{Chao Gong},
  \bibinfo{person}{Yang Shen}, \bibinfo{person}{Jie Zhou},
  \bibinfo{person}{Siming Chen}, \bibinfo{person}{Tao Gui}, \bibinfo{person}{Qi
  Zhang}, {and} \bibinfo{person}{Xuanjing Huang}.}
  \bibinfo{year}{2023}\natexlab{}.
\newblock \bibinfo{title}{A Comprehensive Capability Analysis of GPT-3 and
  GPT-3.5 Series Models}.
\newblock
\newblock
\showeprint[arxiv]{2303.10420}~[cs.CL]


\end{thebibliography}
  
\end{document}